\def\IEEEtemp{}
\ifx\IEEEtemp\undefined
\documentclass[12pt]{article}
\usepackage[T1]{fontenc}\begin{flushright}
	
\end{flushright}
\usepackage{geometry}
\else
\documentclass[10pt,journal,compsoc]{IEEEtran}
\fi

\def\allfiles{main}

\usepackage[linesnumbered,ruled,vlined]{algorithm2e}
\usepackage{algpseudocode}
\usepackage{amsmath,amssymb}
\usepackage{mathrsfs}
\usepackage{amsthm}
\usepackage{bm}
\usepackage{multirow}
\usepackage{booktabs}
\usepackage{graphics}
\usepackage{epstopdf}

\usepackage{epsfig}
\usepackage{footnote}

\usepackage{array}
\usepackage{stfloats}
\usepackage{color}
\usepackage[numbers,sort&compress]{natbib}
\usepackage{bm}
\usepackage{enumitem}
\usepackage{epsfig}

\usepackage{footmisc}

\PassOptionsToPackage{hyphens}{url}\usepackage[hidelinks]{hyperref}
\makeatletter
\def\UrlAlphabet{%
	\do\a\do\b\do\c\do\d\do\e\do\f\do\g\do\h\do\i\do\j%
	\do\k\do\l\do\m\do\n\do\o\do\p\do\q\do\r\do\s\do\t%
	\do\u\do\v\do\w\do\x\do\y\do\z\do\A\do\B\do\C\do\D%
	\do\E\do\F\do\G\do\H\do\I\do\J\do\K\do\L\do\M\do\N%
	\do\O\do\P\do\Q\do\R\do\S\do\T\do\U\do\V\do\W\do\X%
	\do\Y\do\Z}
\def\UrlDigits{\do\1\do\2\do\3\do\4\do\5\do\6\do\7\do\8\do\9\do\0}
\g@addto@macro{\UrlBreaks}{\UrlOrds}
\g@addto@macro{\UrlBreaks}{\UrlAlphabet}
\g@addto@macro{\UrlBreaks}{\UrlDigits}
\makeatother

\newcommand{\cR}{\mathsf{R}}
\newcommand{\Budget}{\mathnormal{\tau}}

\newcommand{\huaA}{\mathcal{A}}
\newcommand{\huaC}{\mathcal{C}}
\newcommand{\huaL}{\mathcal{L}}

\newcommand{\Enc}{\mathsf{Enc}}
\newcommand{\Dec}{\mathsf{Dec}}

\newtheorem{theorem}{Theorem}

\newtheorem{definition}{Definition}

\newcommand{\system}{ZebraLancer}
\newcommand{\sameprefix}{common-prefix}

\newcommand{\cCertGen}{\mathsf{CertGen}}
\newcommand{\cCredGen}{\mathsf{CredGen}}

\newcommand{\cCertVrfy}{\mathsf{CertVrfy}}

\newcommand{\cSetup}{\mathsf{Setup}}
\newcommand{\cParam}{\mathsf{Param}}
\newcommand{\Auth}{\mathsf{Auth}}
\newcommand{\Verify}{\mathsf{Verify}}
\newcommand{\Link}{\mathsf{Link}}
\newcommand{\cC}{\mathcal{C}}
\newcommand{\Challenge}{\mathsf{Challenge}}

\newcommand{\cA}{\mathcal{A}}

\newcommand{\Contract}{\mathscr{C}}

\newcommand{\cProve}{\mathsf{Prover}}
\newcommand{\cVerify}{\mathsf{Verifier}}

\newcommand{\ignore}[1]{}

\IEEEoverridecommandlockouts

\begin{document}

\title{\system: Decentralized Crowdsourcing of Human Knowledge atop Open Blockchain}

\author{Yuan~Lu,
	Qiang~Tang 
	and~Guiling~Wang
	\IEEEcompsocitemizethanks{\IEEEcompsocthanksitem The authors are with the Department
		of Computer Science, Ying Wu College of Computing, New Jersey Institute of Technology, Newark,
		NJ, 07102, USA.	E-mail: \{yl768, qiang, gwang\}@njit.edu
		\IEEEcompsocthanksitem  Two popular hypotheses of zebra strips were: (i) camouflage used to	confuse predators by motion dazzle, and (ii) visual cues used by herd peers to identify. The delicate anonymity in our system can be analog, as it overcomes the natural tension between anonymity and accountability.}
	\thanks{Manuscript received Dec xx, 2018. An abridged version of this paper appeared in the 38th IEEE ICDCS, Vienna, Austria, July 2018.}}

\markboth{IEEE TRANSACTIONS ON XXXX,~Vol.~xx, No.~xx, Xxx~20xx}%
{Shell \MakeLowercase{\textit{et al.}}: Bare Demo of IEEEtran.cls for Computer Society Journals}

\IEEEtitleabstractindextext{%
	\begin{abstract}
		We design and implement the first {\em private and anonymous} decentralized crowdsourcing system \system\footnotemark{}, and overcome two fundamental challenges of decentralizing crowdsourcing,  i.e., data leakage and identity breach.
				
		First, our {\em outsource-then-prove} methodology resolves the tension between the blockchain transparency and the data confidentiality to guarantee the basic utilities/fairness requirements of data crowdsourcing, thus ensuring: (i) a requester will not pay more than what data deserve, according to a policy announced when her task is published via the blockchain; (ii) each worker indeed gets a payment based on the policy, if he submits data to the blockchain; (iii) the above properties are realized not only without a central arbiter, but also without leaking the data to the open blockchain.
		Second, the transparency of blockchain allows one to infer private information about workers and requesters through their participation history. Simply enabling anonymity is seemingly attempting but will allow malicious workers to submit multiple times to reap rewards. \system{\ }also overcomes this problem by allowing anonymous requests/submissions without sacrificing the accountability. The idea behind is a subtle linkability: if a worker submits twice to a task, anyone can link the submissions, or else he stays anonymous and unlinkable across tasks. To realize this delicate linkability, we put forward a novel cryptographic concept, i.e., the {\em common-prefix-linkable} anonymous authentication.
		We remark the new anonymous authentication scheme might be of independent interest.
		Finally, we implement our protocol for a common image annotation task and deploy it in a test net of Ethereum. The experiment results show the applicability of our protocol atop the existing real-world blockchain.
	\end{abstract}
	
	}

\maketitle
\IEEEdisplaynontitleabstractindextext
\IEEEpeerreviewmaketitle


\IEEEraisesectionheading{\section{Introduction}\label{introduction}}

Crowdsourcing empowers open collaboration over the Internet. One remarkable example is the solicitation of annotated data: the benchmark of famous ImageNet challenge \cite{DDS09} was created via Amazon's crowdsourcing marketplace, Mechanical Turk (MTurk) \cite{MTurk}. Another notable example is mobile crowdsensing \cite{GYL11} where one (called ``requester'') can request a group of individuals (called ``workers'') to use mobile devices to gather information fostering data-driven applications \cite{WazeDown,DSL13}.
%
Various monetary incentive mechanisms were introduced  \cite{WSZ15,ZV12,ZLM14,YXF12,PWC15,SZ15,JSC15} to motivate workers to make real efforts.
To facilitate these mechanisms, the state-of-the-art solution necessarily requires a trusted third-party to host crowdsourcing tasks to fulfill the fair exchange between the crowd-shared data and the rewards
; otherwise, the effectiveness of incentive mechanisms can be hindered by the so-called ``free-riding'' (i.e. dishonest workers reap rewards without making real efforts) and ``false-reporting'' (i.e. dishonest requesters try to repudiate the payment).

\smallskip
It is well-known that the reduction of the reliance on a trusted third-party is desirable in practice, and the same goes for the case of crowdsourcing.
First, numerous real-world incidents reveal that the party might silently misbehave in-house for self-interests \cite{Qihoo}; or, some of its employees \cite{Alipay} or  attackers \cite{Apple} can compromise its functionality. Second, the party often fails to resolve disputes. For instance, requesters have a good chance to collect data without paying at MTurk, because the platform is biased on requesters over workers \cite{MCN16}. Third, a centralized platform inevitably inherits all the vulnerabilities of the single point failure. For example, Waze, a crowdsourcing map app, suffered from 3 unexpected server downs and 11 scheduled service outages during 2010-2013 \cite{WazeDown}. Last but not least, a central platform hosting all tasks also increases the worry of massive privacy breach. 
A most fresh lesson to us is the tremendous data leakage of Uber \cite{Uber}.

\smallskip
In contrast, an open blockchain\footnotemark{ }is a distributed, transparent and immutable ``bulletin board'' organized as a chain of blocks. It is usually managed and replicated by a peer-to-peer (P2P) network collectively. Each block will include some messages committed by network peers, and be validated by the whole network according to a pre-defined consensus protocol. This ensures reliable message deliveries via the untrusted Internet. More interestingly, the messages contained in each block can be program code, the execution of which is enforced and verified by all P2P network peers; hence, a more exotic application of smart contract \cite{Sza97} is enabled. Essentially, the smart contract  can be viewed as a ``decentralized computer'' that faithfully handles all computations and message deliveries related to a specified task. It becomes enticing to build a {\em decentralized} crowdsourcing platform atop.

\footnotetext{ We remark that blockchain is used to refer open blockchain through this paper. An open/permissionless/public blockchain is a blockchain network that allows any party to participate and leave, as opposed to a less ambitious way of building blockchain atop permissioned parties. \system{ }can inherit the P2P network of open blockchain as the underlying infrastructure.}

\smallskip
Unfortunately, this new fascinating technology also brings about new privacy challenges \cite{KateTalk}, which were never that severe in the centralized setting before and could even harm the basic system utilities, as one notable feature of the blockchain is its {\em transparency}. The whole chain is replicated to the whole network to ensure consistency, thus the data submitted to the blockchain will be visible to the public. 

\smallskip
First, the blockchain transparency causes an immediate problem violating data privacy, considering that many of the crowdsourced data maybe sensitive. For example, even in the intuitively ``safe'' image annotation tasks, if there are some special ambiguous pictures (e.g. Thematic Apperception Test pictures \cite{TAT}), the answers to them can be used to infer the personality profiles of workers. Sometimes, the data are simply valuable to the requester who paid to get them. What is worse, since the block confirmation (which corresponds to the time when  the submitted answers are actually recorded in a block) normally takes some time after the data is submitted to the network, a malicious worker can simply copy the data committed by others, and submit the same data as his own to run the free-riding attack. Without {\em data confidentiality}, the incentive mechanisms could be rendered completely ineffective in decentralized settings. 

\smallskip
Second, most crowdsourcing systems \cite{MTurk,WazeDown} and incentive mechanisms \cite{ZV12,ZLM14,YXF12} implicitly require requesters/workers to authenticate to prevent misbehaviors caused by (colluded) counterfeited identities \cite{Dou02}. When  crowdsourcing is decentralized, this basic requirement will cause the history of submitting/requesting to be known by everyone (which was previously ``protected'' in a data center such as the breached one of Uber's). Thus considerable information about workers/requesters \cite{YZR15} will leak to the {\em public} through their participation history, which seriously impairs their privacy and even demotivates them to join.
Notably, if a user frequently joins traffic monitoring tasks, 
then {\em anyone} can read the blockchain and figure out his location traces.




\smallskip
Clearly, to decentralize crowdsourcing in a meaningful way to realize basic system utilities, the above fundamental privacy challenges have to be overcome, which requires us to resolve two natural tensions: (i) the tension between the blockchain transparency and the data confidentiality, and (ii) the tension between the anonymity and the accountability. Simple solutions utilizing some standard cryptographic tools (e.g. encryption and/or group signature) to protect the data confidentiality and the anonymity do not work well: the encryption of data immediately prevents smart contracts from enforcing the rewards policy; to allow fully anonymous participation will give a dishonest worker an opportunity of multiple submissions in one crowdsourcing task, and thus he may claim more rewards than what is supposed (similarly for a malicious requester).
See more details about the challenges in Section \ref{problem}.

\smallskip
\noindent{\bf Our contributions.} 
In this paper, we construct, analyze and implement a general blockchain based protocol to enable the first {\em private and anonymous} decentralized data crowdsourcing system\footnote {A couple of recent attempts on decentralized crowdsourcing \cite{LWY17} have been made, however, none resolved the fundamental privacy issues, which are quite arguable for basic functionalities, as in these systems, malicious workers can simply free-ride by copying other submissions. See Section \ref{related} for the detailed insufficiencies.}. 
Without relying on any third-party information arbiter, our protocol can still guarantee the faithful execution for a class of incentive mechanisms, once they are announced as pre-specified policies in the blockchain.
More importantly, we also protect confidential data and identities from the blockchain network, while the underlying blockchain is auditing the correct execution of crowdsourcing tasks. Specifically,

\begin{enumerate}
\smallskip
\item  
    A blockchain based protocol is proposed to realize decentralized crowdsourcing that satisfies: (i) the fair exchange between data and rewards, i.e., a worker will be paid the correct amount according to the pre-defined policy of evaluating data, if he submits to the blockchain; (ii) data confidentiality, i.e., the submitted data is confidential to anyone other than the requester; and (iii) anonymity and accountability.  

Intuition behind the fairness and confidentiality is an {\em outsource-then-prove} methodology that: (i) the requester is enforced to deposit the budget of her incentive policy to a smart contract; (ii) submissions are encrypted under the requester's public key and will be collected by the blockchain; (iii) the evaluation of rewards is {\em outsourced} to the requester who then needs to send an instruction about how to reward workers. The instruction is ensured to follow the promised incentive policy, because the requester is also required to attach a valid succinct zero-knowledge proof.

The worker anonymity of our protocol can ensure: (i) the public, including the requester and the implicit registration authority, is not able to tell whether a data comes from a given worker or not; (ii) if a worker joins multiple tasks announced via the blockchain, no one can link these tasks. More importantly,
we also address the threat of multiple-submission exacerbated by anonymity misuse. In particular,  if a worker anonymously submits more than the number allowed in {\em one} task, our scheme allows the blockchain to tell and drop these invalid submissions. Similarly, we achieve the requesters' accountable anonymity.

\ignore{
The intuition behind our solution is that we augment a non-interactive procedure to be a two-step interactive protocol: at beginning, the requester announces a data crowdsourcing contract that specifies the task and the quality condition for the payment; in the first round, workers submit data encrypted under requester's public key to the blockchain. Since the contract now cannot process ciphertext, we ``outsource'' the decryption procedure to the requester offline and let the requester to instruct the contract how to pay; then the contract will automate the payments. To ensure the (potentially dishonest) requester to send back a proper instruction, in the second round we require the requester to send his instruction together with an succinct zero-knowledge proof.
The proof has to convince the contract that the requester indeed crafts the instruction as specified. Note that if a malicious requester tries to do false-reporting, the proof cannot be verified, thus the contract will not be executed further steps. This now becomes not of the requester's interests as all the requester's rewards were deposited/locked in the contract, a false-reporting makes the requester lose more. While free-riding cannot happen as the workers can only get paid after the requester checks the data quality (after decryption).
}

\smallskip
\item  
To achieve the above goal of anonymity while preserving accountability, we propose, define and construct a new cryptographic primitive, called {\em \sameprefix\/-linkable} anonymous authentication. In most of the time, a user can authenticate messages and attest the validity of identity without being linked. The only exception that anyone can link two authenticated messages is that they share the same prefix and are authenticated by one user.

To utilize the new primitive in our protocol, a worker has to submit to a task via an anonymous authentication. The reference of the task will be unique, and should be the common-prefix, so that the special linkability will prevent multi-submission to a task. A requester can also use it to authenticate in each task she publishes, and convince workers that she cannot maliciously submit to intentionally downgrade their rewards. We remark that such a scheme can be used to anonymously authenticate  in a constant time independent to tasks.

Such a primitive may be of independent interests for the special flavor of its accountability-yet-anonymity.

\smallskip
\item  
To showcase the feasibility of applying our protocol, we implement the system that we call ZebraLancer for a common image annotation task on top of Ethereum, a real-world blockchain infrastructure. Intensive experiments and performance evaluations are conducted in a Ethereum test net. Since current smart contracts support only  primitive operations, tailoring such protocols compatible with existing blockchain platforms is non-trivial.
\end{enumerate}


\ifx\allfiles\undefined
\bibliographystyle{IEEEtran}
\bibliography{reference}

\end{document}
\fi

\medskip


\section{Problem Formulation}\label{problem}

In this section, we will give more precise definitions about the problem and its security requirements.

\begin{table}[!htbp]
	\renewcommand{\arraystretch}{1.3}
	\caption{Key notations related to crowdsourcing model}
	\label{table:problem-notation}
	\begin{tabular}{m{1.2cm}||m{6.5cm}}
		\hline
		Notation & Represent  for \\
		\hline\hline
		$R$ & A requester that can be uniquely identified by ID $R$                  	\\[2pt]
		$W_j$ & A worker who can be uniquely identified by ID $W_i$                  	\\[2pt]
		$T$ & The crowdsourcing task published by $R$ to collect $n$ answers from $n$ different workers                 	\\    		[2pt]
		$A_j$ & The answer that is submitted  to $T$ by $W_j$                  	\\[2pt]
		$\cR(A_j;\tau)$ & The process that defines the value of the answer $A_j$                  	\\[2pt]
		$R_j$ & The reward for answer $A_j$ according to its value              	\\[2pt]
		\hline
	\end{tabular}
\end{table}

\begin{figure}[!h]
\centering
\includegraphics[width=8.5cm]{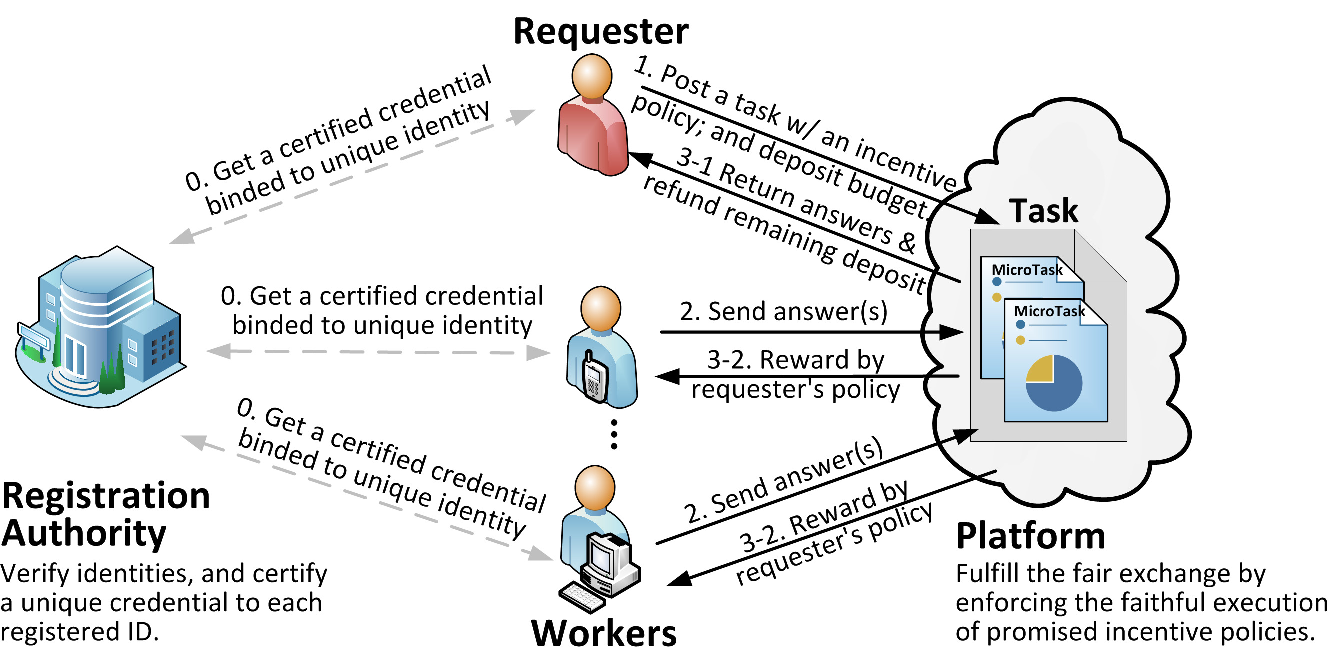}
\caption{The model of data crowdsourcing: workers and requesters obtain unique credentials bound to their identities at a registration authority (RA); authenticated requesters and authenticated workers can make fair exchange between data and rewards through a third-party platform (arbiter).}
\label{fig:model}
\end{figure}

\smallskip
\noindent{\bf Data crowdsourcing model.}
\label{sec2-1}
%
As illustrated in Fig.\ref{fig:model}, there are four roles in the model of data crowdsourcing, i.e., requesters, workers, a platform and a registration authority. A \emph{requester}, uniquely identified by $R$, can post a task to collect a certain amount of answers from the crowd. When announcing the task, the requester promises a concrete reward policy to incentivize workers to contribute (see details about the definition of reward policy below). A \emph{worker} with a unique ID $W_j$, submits his answer $A_j$ and expects to receive the corresponding reward. The \emph{platform}, a medium assisting the exchange between requesters and workers, is either a trusted party or emulated by a network of peers. The platform considered in this paper is jointly maintained by a collection of network peers, and in particular, we will build it atop a open blockchain network. The \emph{registration authority} (RA),
can play an important role of verifying and managing unique identities of workers/requesters, by binding each identity to a unique credential (e.g. a digital certificate). Note that an answer $A_j$ is chosen from a well-defined set $\bm A$.


\smallskip
We remark that the well established identities are necessary demand of real-world crowdsourcing systems, for example MTurk and Waze, to prevent misbehaviors such as Sybil attack. Moreover, many auction-based incentive mechanisms \cite{ZLM14,YXF12} are built upon the non-collusive game theory that implicitly requires established identities to ensure one bid from one unique ID. We employ RA to establish identities. In practice, a RA can be instantiated by (i) the platform itself, (ii) the certificate authority who provides authentication service, or (iii) the hardware manufacturer who makes trusted devices that can faithfully sign messages \cite{SW10}. Our solution should be able to inherit these established RAs in the real-world.

\smallskip
In this paper, w.l.o.g., we assume that each unique identity is only allowed to submit one answer to a task. Also, we consider settings where the value of crowd-shared answers can be evaluated by a well-defined process such as auctions and quality-aware rewards, and also the corresponding rewards, c.f. \cite{PWC15,SZ15,JSC15} about quality-aware rewards and \cite{ZLM14,YXF12} about auction-based incentives. These incentives share the same essence as follows.

\smallskip
Suppose an authenticated requester publishes a task $T$ with a budget $\Budget$ to collect $n$ answers from  $n$ workers. An authenticated worker interested in it will then submit his answers.
%
The {\em reward} of an answer $A_j$ will follow a well-defined process determined by some auxiliary  variables,
i.e., the reward of $A_j$ can be defined as $R_j:=\cR(A_j; a_1 , \ldots , a_m, \tau)$, where $\cR$ is  a function parameterized by some auxiliary variables denoted by $a_1,\ldots,a_m, \tau$. Remark that $\tau$ is the budget of the requester, and we will use $R_j:=\cR(A_j;\tau)$ for short.

\smallskip
Particularly, in some simple tasks (e.g. multiple choice problems), the quality of an answer can be straightforwardly evaluated by all answers to the same task, i.e.  $R_j=\cR(A_j; A_1 , \ldots , A_n, \tau)$, with using majority voting or estimation maximization iterations \cite{PWC15,SZ15,JSC15}. More generally, \cite{BK13} proposed a universal method to evaluate quality:
(i) some workers are requested to answer a complex task; (ii) other different workers are then requested to grade each answer collected in the previous stage.
What's more, our model captures the essence of auction-based incentive mechanism such as \cite{YXF12,ZLM14}, when the parameters $a_1 , \ldots , a_m$ represent the bids of workers (and other necessary auxiliary inputs such as their reputation scores). 

\smallskip
Our work considers the general definition above and will be extendable to the scope of auction-based incentives, even though the protocol design and implementations in this paper mainly focus on how to instantiate quality-aware incentives. Also note that the flat-rate incentive, that is each submitted answer will get a flat-rate payment \cite{MTurk}, can be considered as a special case of the above abstraction as well.



%



\smallskip
\noindent{\bf Security models.}
\label{sec2-3}
Next, we specify the basic security requirements for our (decentralized) crowdsourcing system on top of the existing infrastructure of open blockchain.


\smallskip
{\em Data confidentiality}. Ideally, this property requires that the communication transcripts (including the blocks in the blockchain) do not leak anything to anyone (except the requester) about the input parameters $a_1 , \ldots , a_m$ of the incentive policy $\cR$. Because these parameters might actually be confidential data or sealed bids. We can adapt the classical semantic security \cite{GM82} style definition from cryptography for this ideal purpose: the distribution of the public communication can be simulated with only public knowledge. 
However, an adversarial worker can infer information after receiving his payment. This is the inherent issue of incentive mechanisms, and cannot be resolved even if there is a fully trusted third-party. So our protocol will focus on the best-possible data confidentiality, as if there is a fully trusted party. Informally speaking, our protocol is running in the ``real'' world, and imagine there is an ``ideal'' world where exists a fully trusted party facilitating the incentives; by real-ideal world paradigm, our data confidentiality requires that the two worlds are computationally indistinguishable.

\smallskip
{\em Anonymity}. Private information of worker/requester can be explored by linking tasks they join/publish \cite{YZR15}. Intuitively, we might require two anonymity properties for workers: (i) {\em unlinkability between a submission and a particular worker} and (ii) {\em unlinkability among all tasks joined by a particular worker}. However, (i) indeed can be implied by (ii), because the break of first one can obviously lead up to the break of the latter one. Similarly, the anonymity of requester can be understood as the unlinkability among all tasks published by her.
The requirement of worker anonymity can be formulated via the following game. An adversary $\huaA$ corrupts a requester, the registration authority (RA), and the platform (e.g. the blockchain); suppose there are only two honest workers, $W_0$ and $W_1$. In the beginning, the adversary announces a number of $n$ tasks. For each task $T_i$, suppose there are a set of participating workers $\bm{W}_{T_i}$. After seeing all the communications, for any $T_i\neq T_j$, $\huaA$ cannot tell whether $\bm{W}_{T_i} \cap \bm{W}_{T_j} =\emptyset$ better than guessing. We note that the anonymity should hold even if all entities, including the requester and the platform (except $W_0$ and $W_1$), are corrupted.
The requester anonymity can be defined via the above game similarly, and we omit the details.


\smallskip
{\em Security against a malicious requester.} A malicious requester may avoid paying rewards (defined by the policy $\cR$), e.g. launches the {\em false-reporting} attack.
Security in this case can be formulated via the following security game: an adversary $\huaA$ corrupts the requester and executes the protocol to publish a task with a promised reward policy $\cR$ and a budget $\tau$.
Let us define a bad event $B_1$ to be that there exists a worker $W_j$, who submits answers $A_j$ and receives a payment smaller than $R_j=\cR(A_j;\tau)$. We require that for every polynomial time adversary $\huaA$, the probability $\Pr[B_1]$ is negligibly small.




\smallskip
{\em Security against malicious workers.} A dishonest worker may try to harvest more rewards than what he deserves. Security in this case can be formulated as follows. An adversary $\huaA$ corrupts one worker,\footnote{We remark that we focus on resolving the {\em new} challenges introduced by blockchain, and put forth the best possible security,  as if there is a fully trusted third-party serving as the crowdsourcing platform. For example, it is not quite clear how to handle the collusion of many identities, even in the centralized setting; thus such a problem is out of the scope of this paper.}
and participates in the protocol interacting with a requester (and the platform): (i) $\huaA$ submits some answers $\{A_1,\ldots,A_n\},n \geq 1$,
let us define the bad event $B_2$ as that $\huaA$ receives a payment greater than $\max_{\{A_j \in \{A_1,\ldots,A_n\} \}} R_j:=\cR(A_j;\Budget)$ from the requester; 
(ii) $\huaA$ also sees the transcripts corresponding to an honest worker $W_i$'s submission (e.g. the ciphertext of $A_i$), and then submits her answer $A_j$ to the platform, let us define the bad event $B_3$ as that $\huaA$ receives a payment $R_j = \cR(A_i;\Budget)$.
We require that for all polynomial time $\huaA$, $\Pr[B_2] $ and $\Pr[B_3]- \Pr[ (A_\huaA \leftarrow \bm{A}) = A_i ]$ are negligibly small.
%

\smallskip
We remark that the above securities against a malicious requester and malicious workers have captured the special fairness of the exchange between crowd-shared data and rewards. For example, Sybil attackers who forge identities, and front-running attackers who copy and paste others' submissions (which can be either ciphertexts or plaintexts) shall be prevented.

\ignore{
	Remark that the above definition has captured the free-riding and (uncolluded) Sybil attacks as special cases.
	For example, a Sybil attacker who corrupts a worker and tries to send two answers ($n=2$) to get extra payment. We further remark that we only focus on how to enforce incentive mechanisms, and achieve a best possible security, as if there was a trusted party. In case there are more sophisticated mechanisms to further de-incentivize colluded workers, we might apply them in our system by instantiating concrete $\cR$.
}


\medskip
\noindent{\bf Technical challenges.} The main advantage that the blockchain offers is the smart contract that can be automatically executed as a piece of pre-defined agreement.
Let us firstly look at a {\em naive} decentralized solution to a crowdsourcing task: the requester codes her incentive policy into a task's smart contract $\Contract$, and then broadcasts the contract $\Contract$ to the blockchain network to collect $n$ answers; the incentive policy is using an incentive mechanism defined by a function $\cR$ and a budget $\Budget$; after the contract $\Contract$ is included by a block, it can expect workers submit answers to it in the clear; finally, $\Contract$ can evaluate the reward $R_i=\cR(A_i; A_1 , \ldots , A_, \tau)$ for each answer $A_i$; more importantly, the contract can further automatically transfer $R_i$ to the blockchain address bound to answer $A_i$.

\medskip
The above naive solution looks appealing to resolve the fairness issues such as {\em free-riding} and {\em false-reporting}.
Notice that an implicit requirement is that answers (or bids) should be submitted in the clear, as the contract needs all those as the inputs (i.e. $a_1 , \ldots , a_m$) of incentive policy $\cR$ to audit the right amount of the reward for each answer. However, those inputs can be valuable and sensitive crowd-shared answers, and should be kept confidential. Or they could be auction bids, and should be confidential as well, because a malicious worker can learn the distribution of truthful bids of honest workers, and further break auctions by crafting optimized malicious (untruthful) bids. One may suggest that a worker encrypts his answer (or his bid) under the requester's public key, and submits the ciphertext instead. Unfortunately, this immediately renders the above naive solution to fail, as the contract (more precisely, the blockchain peer) only sees ciphertexts and thus cannot evaluate the reward. Another proposal to hardcode the secret key in the smart contract also fails, as the secret key will become transparent. Also remark that more advanced encryption schemes such as fully homomorphic encryption do not help, because the blockchain peers still cannot know the plaintext of rewards without the secret key, even if they can compute the ciphertext of rewards. Therefore, the tension between the data confidentiality and the transparent execution of smart contracts brings in our first challenge:

\textbf{Challenge 1}. \emph{Leveraging the blockchain to enforce incentives, but not revealing the confidential inputs of incentive mechanism (e.g. answers or bids) to the public.}

\medskip
As briefly pointed out in the introduction, when the anonymity of workers/requesters is not protected well, their private information can be hampered, because the participation history of tasks will leak to the public through the blockchain. What's worse, negative payoffs will be added to decrease the motivations of workers/requesters to participate, due to their breached privacy.
To mitigate the risk of privacy breach, one might suggest anonymous submission/request.
However, if a worker is allowed to submit data without authenticating his unique identity, a malicious one may exploit this to flood a task by multiple submissions. In particular, a requester specifies a task to collect 20 answers from 20 anonymous workers, and a malicious worker might submit 20 colluded answers with claiming that they are from 20 different workers, as all answers are anonymously submitted. Worse still, an anonymous requester might send (colluded) answers to her own task to repudiate payments. For instance, a requester anonymously publishes a task to collect 50 answers, and she submits 30 colluded anonymous answers to downgrade other 20 answers, such that she avoids paying but still gets 20 useful answers. These scenarios cause another tension between the participants anonymity and their accountability, which brings in the second challenge:


\textbf{Challenge 2}. \emph{Ensuring participants to be anonymous while keeping their accountability in decentralized crowdsourcing.}

\medskip

\section{Related Work}\label{related}
We thoroughly review related works, and briefly discuss the insufficiencies of the state-of-the-art solutions.

\smallskip
\noindent{\bf Centralized crowdsourcing systems}.
MTurk~\cite{MTurk} is the most commercially successful crowdsourcing platform. But it
has a well-know vulnerability allowing false-reporters gain short-term advantage \cite{MCN16}. Also, MTurk collects plaintexts of answers, which causes considerable worry of data leakage. Last, the pseudo IDs in MTurks can be trivially linked by a malicious requester. Dynamo \cite{SIB15} was designed as a privacy wrapper of MTurk. Its pseudo ID can only be linked by the pseudo ID issuer, but still it inherited all other weaknesses of MTurk.
SPPEAR \cite{GGP16} considered a couple of privacy issues in data crowdsourcing, and thus introduced a couple more authorities, each of which handled a different functionality. Distributing one authority into multiple reduces the excessive trust, but, unfortunately, it is still not clear how to instantiate all those different authorities in practice. 



\smallskip
\noindent{\bf Decentralized crowdsourcing}.
We also note there are several attempts \cite{BLN17,CSJ15,LWY17,Ant17} using blockchain to decentralize crowdsourcing, but neither of them considers  privacy and anonymity which are arguably fundamental for basic utility:
the authors of \cite{BLN17} built up a crowd-shared service on top of the blockchain, but the system is not compatible with incentive mechanisms and is not privacy-preserving either; the authors of \cite{CSJ15} leveraged the blockchain as a payment channel in their crowdsourcing framework, but it is neither secure against malicious workers and dishonest requesters, nor privacy-preserving; a couple of recent attempts \cite{LWY17,Ant17} took advantage of the public blockchain to enforce incentives, but these frameworks are neither private nor anonymous, i.e., the collected data and the unique identities (such as certificates) will leak to the whole network of the open blockchain.


\smallskip
\noindent{\bf Anonymous crowdsourcing}.
Li and Cao \cite{LC14} proposed a framework to allow workers generate their own pseudonyms based on their device IDs. But the protocol sacrificed the accountability of workers, because workers can forge pseudonyms without attesting that they are associated to real IDs, which gave a malicious worker chances to forge fake pseudo IDs and cheat for rewards.
Rahaman et al. \cite{RCY17} proposed an anonymous-yet-accountable protocol for crowdsourcing  based on group signature, and focused on how to revoke the anonymity of misbehaved workers. Misbehaved workers could be identified and further revoked by the group manager. The authors in \cite{GGP16} similarly relied on group signature but introduced a couple of separate authorities.
Our solution can be considered as a proactive version that can prevent worker misbehavior, and without relying on a group manager. 

\smallskip
\noindent{\bf Accountable anonymous authentication}.
%
The pioneering works in anonymous e-cash \cite{Cha83,CFN90} firstly proposed the  notion of one-time anonymous authentication. The concept later was studied in the context of one-show anonymous credential \cite{CL01}. 
%
Some works \cite{XY04,TFS04} further extended the notion of one-time use to be {\em k}-time use, and therefore enabled a more general accountability for anonymous authentications.
In \cite{CHK06}, the authors considered a special flavor of accountability to periodically allow {\em k}-time anonymous authentications.

Our new primitive provides a more fine-grained conditional linkage of anonymous authentications, which is needed for preventing multi-submission in each crowdsourcing task.


\smallskip
\noindent{\bf Linkable group/ring signature}. 
Conceptually similar to the linkability appeared in one-show credential \cite{CL01}, linkable group/ring signatures \cite{NFW99,LWW04} were proposed to allow a user to sign messages on behalf of his group unlinkably up to twice.
 
In \cite{ASY06}, a more general concept of event-oriented linkable group signature was formulated to realize more fine-grained trade-off between accountability and anonymity: a user can sign on behalf of his group unlinkably up to $k$ times per $event$, where an $event$ could be a common reference string (e.g., the unique address to call a smart contract deployed in the blockchain). But its main disadvantage is that the group manager can reveal the actual identities of users. Similar event-oriented linkability was discussed in the context of ring signature \cite{TWC04} as well. Even though that construction enjoys unbreakable anonymity, its main drawback becomes the impracticability of ``hiding'' the real identity behind a large group (mainly because the verification of signatures might require the public keys of all group members).

Our new primitive can be considered as a special cryptographic notion to formalize the subtle balance between event-oriented linkability and irrevocable anonymity (within a large and dynamic group). Specifically, our scheme ensures that no one can link the actual identity to any authenticated message (which is strictly stronger than \cite{ASY06}), and also it enables a user to ``hide'' behind a large group of users (i.e. all users registered at RA, which is impossible in practice via \cite{TWC04}).

\smallskip
\noindent{\bf Privacy-preserving smart contracts}.
Privacy-preserving smart contract is a recent hot topic in blockchain research. Most of them are for general purpose consideration \cite{KZZ16,ZNP15}, and thus deploy heavy cryptographic tools including general secure multi-party computation (MPC). Hawk \cite{KMS16} did provide a general framework for privacy-preserving smart contracts using light zk-SNARK, but mainly for reward receiver to prove to the contract. 
Our work can be considered as a very specially designed MPC protocol, and a lot of dedicated optimizations of zk-SNARK exist which can directly benefit our protocol. Last, cryptocurrencies like Zcash \cite{HBH17} and Ethereum \cite{Byzantium} also leverage zk-SNARK to build a public ledger that supports anonymous transactions. We note that they consider more basic blockchain infrastructures, on top of which we may build our application for crowdsourcing.

\medskip
\ifx\allfiles\undefined
\documentclass[12pt]{article}
\usepackage{comment}
\usepackage[linesnumbered,ruled,vlined]{algorithm2e}
\usepackage{algpseudocode}
\usepackage{amsmath,amssymb}
\usepackage{mathrsfs}
\usepackage[hyphens]{url}
\usepackage[T1]{fontenc}
\usepackage{geometry}
\geometry{left=2.5cm,right=2.5cm,top=2.5cm,bottom=2.5cm}
\linespread{1.6}
\usepackage{graphicx}
\begin{document}
\setcounter{section}{1}
\fi


\section{Preliminaries}\label{preliminary}


\noindent\textbf{Blockchain and smart contracts.} A blockchain is a global ledger maintained by a P2P network collectively following a pre-defined consensus protocol. 
Each block in the chain will aggregate some transactions containing use-case specific data (e.g., monetary transfers, or program codes).
%
%
In general, we can view the blockchain as an ideal public ledger \cite{GKL15} where one can write and read data in the clear. Moreover, it will faithfully carry out certain pre-defined functionalities. The last property  captures the essence of smart contracts, that every blockchain node will run them as programs and update their local replicas according to the execution results, collectively. More specifically, the properties of the blockchain can be informally abstracted as the following ideal public ledger model \cite{KMS16}:
%
\begin{enumerate}

\smallskip
\item {{\em Reliable delivery of messages}.} The blockchain can be modeled as an ideal public ledger that ensures the {\em liveness and persistence} of messages committed to it \cite{GKL15}. Detailedly, a message sent to the blockchain is in the form of a validly signed transaction broadcasted to the whole blockchain network, and then it will be solicited by a block and written into the blockchain.

Remark that we assume the synchronous network model \cite{KMS16,PSS17}  of blockchain through the paper, that means a transaction will appear in any honest node's replica within a certain time after it appears in the blockchain network. Essentially, the blockchain delivers any message in the form of valid transaction to any blockchain node within a-priori delay.

Also we remark that a network adversary can reorder transactions that are broadcasted to the network but not yet written into a block.

%
\smallskip
\item {{\em Correct computation}.} The blockchain can be seen as a state machine driven by messages included in each block \cite{Vit14}.
Specifically, miners and full nodes will persistently receive newly proposed blocks, and faithfully execute ``programs'' defined by current states with taking messages in new blocks as inputs. Moreover, the computing results can be reliably delivered to the whole network.

\smallskip
\item {{\em Transparency}.} All internal states of the blockchain will be visible to the whole blockchain (intuitively, anyone). Therefore, all message deliveries and computations via the blockchain are in the clear.

\smallskip
\item {{\em Blockchain address (Pseudonym)}.} By default, the sender of a message in the blockchain is referred to a pseudonym, a.k.a. blockchain address. In practice, a blockchain address is usually bound to the hash of a public key; more importantly, the security of digital signatures can further ensure that one cannot send messages in the name of a blockchain address, unless she has the corresponding secret key.

Also, the program code of a smart contract deployed in the blockchain can also be referred by a unique address, such that one can call the contract to be executed, by committing a message pointing to this unique address.




\end{enumerate}

\smallskip
\noindent\textbf{zk-SNARK.}
A zero-knowledge proof (zk-proof) allows a party (i.e. prover) to generate a cryptographic proof convincing another party (i.e. verifier) that some values are obtained by faithfully executing a pre-defined computation on some private inputs (i.e. witness) without revealing any information about the private state. The security guarantees are: 
(i) {\em soundness}, that no prover can convince a verifier if she did not compute the results correctly;
sometimes, we require a stronger soundness that for any prover, there exists an extractor algorithm which interacts with the prover and can actually output the witness (a.k.a. {\em proof-of-knowledge});
(ii) {\em zero-knowledge}, that the proof distribution can be simulated without seeing any secret state, i.e., it leaks nothing about the witness. Both above will hold with an overwhelming probability.

\smallskip
The zero-knowledge succinct non-interactive argument of knowledge (zk-SNARK) further allows such a proof to be generated non-interactively. More importantly, the proof is {\em succinct}, i.e., the proof size is independent on the complexity of the statement to be proved, and is always a small constant. More precisely, zk-SNARK is a tuple of three algorithms. A setup algorithm can output the public parameters to establish a SNARK for a NP-complete language $\huaL = \{ \vec{x} \mid \exists \vec{w},  s.t.,  C(\vec{x},\vec{w})=1 \}$. The $\cProve$ algorithm  can leverage the established SNARK to generate a constant-size proof attesting the trueness of a statement
  $\vec{x}\in\huaL$ with witness $\vec{w}$. The  $\cVerify$  algorithm can efficiently check the proof.




\ifx\allfiles\undefined
\bibliographystyle{IEEEtran}
\bibliography{reference}

\end{document}
\fi

\medskip

\section{Protocol: Private and Anonymous Decentralized Data Crowdsourcing}
\label{sec:freelancer}

In this section, we will construct a private and anonymous protocol to address the critical challenges of decentralizing crowdsourcing, without sacrificing security against ``free-riders'' and ``false-reporters''. The procedures of crowdsourcing will be decentralized atop an existing network of blockchain. More specifically, we will tackle the new privacy and anonymity challenges brought by the blockchain.

\smallskip
As we briefly mentioned in previous sections, the system will implicitly has a separate registration service that validates each participant's unique identity before issuing a certificate. Such setup alleviates some basic problems that every worker is allowed to submit  no more than a fixed number $k$ of answers. For simplicity, we consider here $k=1$.



\smallskip
\noindent{\bf Intuitions.} Our basic strategy is to let the smart contract (which is hosted by the blockchain network) to enforce the fair exchange between the submitted answers and their corresponding rewards,  but without revealing any data or any identity to the blockchain. Let us walk through the high level ideas first.

\smallskip
The requester firstly codifies a reward policy parameterized by her budget (i.e. $\cR(\cdot;\tau)$) into a smart contract. She broadcasts a transaction containing the contract code and the budget. Once the smart contract is included in the blockchain, it can be referred by a unique blockchain address, and the budget should be deposited to this address (otherwise, no one would participate).
After that, any worker who is interested in contributing could simply submit his answer to the blockchain, via a transaction pointing to the contract's address. 

\smallskip
As pointed out before, we have to protect the {\em confidentiality} of the answers, in order to ensure that answers from different workers are independent. So the workers encrypt the answers under the requester's public key.
Now the contract cannot see the answers so it cannot calculate the corresponding rewards. But the requester can retrieve all the encrypted answers and decrypt them off-chain, and further learn the rewards they deserve. It would be necessary that the requester will {\em correctly} instruct the smart contract how to proceed forward.
Concretely, we will leverage the practical cryptographic tool of zk-SNARK to enforce the requester to prove: she indeed {\em followed the pre-specified reward policy} calculating the rewards. Detailedly, the requester should prove her instruction for rewarding is computed as follows: (i) obtain all answers by decrypting all encrypted answers  using a secret key corresponding to the public key contained in the smart contract; (ii) use all those answers and the announced $\cR(\cdot;\tau)$ to compute the quality of each answer.

\smallskip
A more challenging issue arises regarding {\em anonymity-yet-accountability}. Also as briefly pointed before, we would like to
%
%
achieve a balance between anonymity and accountability. Here we put forth a new cryptographic primitive to resolve the  natural tension.
A user can anonymously authenticate on messages (which are composed of a fixed length prefix and the remaining part). But if the two authenticated messages share a common prefix, anyone can tell whether they are done by a same user or not. Moreover, no one can link any two message-authentication pairs, as long as these messages have different prefixes.
 Having this new primitive in hand, a simple and intuitive solution to the anonymous-yet-accountable protocol is to let each worker anonymously authenticate on a message $\alpha_\Contract||C_i$, whenever the encrypted answer $C_i$ is submitted to a task contract $\Contract$ that can be uniquely addressed by $\alpha_\Contract$. This implies that all submissions from the same worker to one task can be linked and then counted, but any two submissions to two different tasks will be provably unlinkable, even if they are submitted by a same user. Also, the number of maximum allowed submissions in each task can be easily tuned (by counting linked submissions).

\smallskip
Last, we also need to augment the smart contract by building the general algorithm of verifying zero-knowledge proofs in it. In particular, when the smart contract receives an instruction regarding rewards and its proof, the verification algorithm will be executed. All inputs of the verification algorithm are common knowledge stored in the open blockchain, e.g., the budget, the encrypted answers and the public key of encryption. If a dishonest requester reports a false instruction, her proof cannot be verified and the contract will simply drop the instruction.
What's more, if the smart contract does not receive a {\em correct} instruction within a time limit, it can directly disseminate the budget to all workers evenly as punishment (which can  be considered as part of the pre-specified incentive mechanism), since the budget has been deposited.
In this way, the requester cannot gain any benefit by deviating from the protocol, and she will be self-enforced to respond properly and timely, resulting in that each worker will receive the expected reward. On the other hand, a dishonest worker can never claim more rewards than that he is supposed to get, as the reward is calculated by the requester herself.

\medskip

    \begin{figure}[!h]
    	\centering
    	\includegraphics[width=8.5cm]{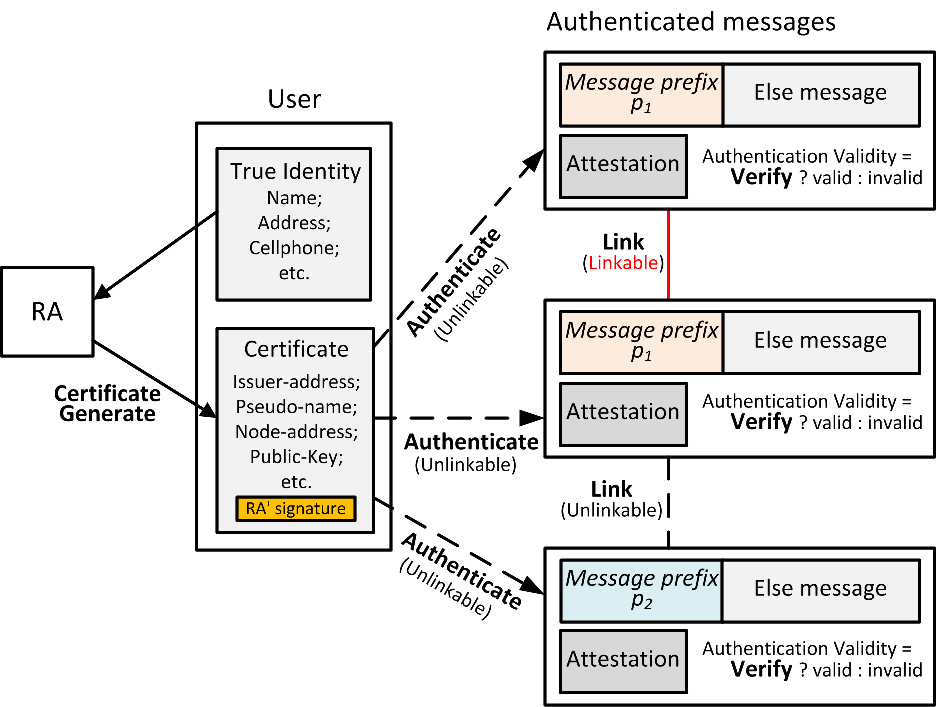}
    	\caption{Subtle linkability of the \sameprefix-linkable anonymous authentication scheme. All involved algorithms except $\cSetup$ are shown in bold.}
    	\label{fig:primitive}
    \end{figure}

  \begin{table}[!htbp]
  		\renewcommand{\arraystretch}{1.3}
  	\caption{Notations related to common-prefix-linkable anonymous authentication}
  	\label{table:primitive-notation}
  	\begin{tabular}{m{1.3cm}||m{6cm}}
  		\hline
  		Notation & Represent  for \\
  		\hline\hline
  		$(mpk,msk)$ & RA's public-secret key pair, which is generated by the algorithm of $\cSetup$                	\\[2pt]
  		$(pk_i,sk_i)$ & The public key certified by RA to bind the unique ID $i$, and its corresponding secret key              	\\[2pt]
  		$\cCertGen$ & The algorithm executed by RA to issue a certificate associated to a registered public key\\[2pt]
  		$\cCertVrfy$ & The algorithm that can be run by anyone to check the validity of a certificate\\[2pt]
  		$\Auth$ & The algorithm to authenticate on a message with using a certified public key\\[2pt]
  		$\Verify$ & The algorithm to verify whether a message is authenticated by a certified public key or not\\	[2pt]
  		$\Link$ & The algorithm to link two authenticated messages, if and only if they share a common-prefix\\	[2pt]
  		$\cProve$ & The zk-SNARK algorithm that generates zk-proofs attesting the statements to be proven              	\\[2pt]
  		$\cVerify$ & The zk-SNARK algorithm that checks whether zk-proofs are generated faithfully or not                	\\[2pt]
  		\hline
  	\end{tabular}
  \end{table}

    \subsection{Common-prefix-linkable anonymous authentication}

    \label{sec:primitive}
     Before the formal description of \system's protocol, let us introduce the new primitive for achieving the anonymous-yet-accountable authentication first. As briefly shown in Fig.\ref{fig:primitive}, our new primitive can be built atop any certification procedure, thus we include a certification generation procedure that can be inherited from any existing one. Also, we insist on {\em non-interactive} authentication, thus all the steps (including the authentication step) are described as algorithms instead of protocols. Formally, a common-prefix-linkable anonymous authentication scheme is composed of the following algorithms:
%

      \begin{enumerate}
      \smallskip \item[-] $\cSetup(1^\lambda)$. This algorithm outputs the system's master public key $mpk$, and system's master secret key $msk$, where $\lambda$ is the security parameter.
      \smallskip \item[-] $\cCertGen(msk,pk)$. This algorithm outputs a certificate $cert$ to validate the public key.
      \smallskip\item[-] $\Auth(m,sk,pk,cert,mpk)$. This algorithm generates an attestation $\pi$ on a message $m$ that: the sender of $m$ indeed owns a secret key corresponding to a valid certificate.
      \smallskip\item[-] $\Verify(m,mpk,\pi)$. This algorithm outputs $0/1$ to decide whether the attestation is valid or not for the attested message, with using system's master public key.
      \smallskip\item[-] $\Link(mpk,m_1,\pi_1,m_2,\pi_2)$. This algorithm takes inputs two valid message-attestation pairs. If $m_1,m_2$ have a common-prefix with length $\lambda$, and $\pi_1,\pi_2$ are generated from a same certificate, it outputs $1$; otherwise outputs $0$.
      \end{enumerate}
      \ignore{
      \begin{enumerate}
      \item[-] $\cSetup(\lambda)$ outputs public parameters $\cParam$. A user can also leverage the established CA as described above, to obtain a certified $PK$ binded to his/her identity.
      \item[-] $\cCredGen(p,sk,pk,cert,mpk;\cParam)$ will allow a user to generate an anonymous credential $cred$ to attest that s/he possesses a secret key paired to a certified public key;  we note that $p$ is a prefix of common knowledge.
      \item[-] $\cCredVrfy(p,cred,PK_{CA};\cParam)$ will allow anyone who receives the anonymous credential $cred$ to get a decision bit $d$, with using CA's public key $PK_{CA}$ and the prefix $p$. If $cred$ is faithfully generated via $\cCredGen$, $d=1$ is expected; otherwise, 0 is expected.
      \end{enumerate}
      }

      We also define the properties for a common-prefix-linkable anonymous authentication.
      {\em Correctness} is straightforward that an honestly generated authentication can be verified. {\em Unforgeability} also follows from the standard notion of authentications, that if one does not own any valid certificate, she cannot authenticate any message. We mainly focus on the formalizing the security notions of {\em \sameprefix-linkability} and {\em anonymity}.

      \ignore{
      \theoremstyle{definition}
      \begin{definition}
      \label{def:correctness}
      \emph{
      	The \emph{correctness} of same-prefix-linkable anonymous credential is: $\forall$ prefix $p$, $\Pr[\cCredVrfy(p,cred,PK_{CA})=1 \mid  pair(PK,SK)=1 \wedge   \cCertVrfy(cert,PK,PK_{CA})=1 \wedge cred=\cCredGen(p,SK,PK,cert,PK_{CA})] = 1 $.
      }
      \end{definition}

      \theoremstyle{definition}
      \begin{definition}
      \label{def:unforge}
      \emph{
      	The \emph{(strong) unforgeability} of same-prefix-linkable anonymous credential can be defined with using the following challenger-adversary game: (i) the challenger generates $PK$ and $SK$, and then registers at CA to obtain a $cert$ binded to $PK$, also the challenger sends $PK$ and $cert$ to the adversary $\huaA$; (ii) the adversary $\huaA$ queries the challenger with a chosen prefix $p_i$, and the challenger computes and sends back $cred_i =\cCredGen(p_i,SK,PK,cert,PK_{CA})$, which will be done for $n$ times; (iii) finally, the adversary $\huaA$ generates a pair of prefix and credential $(p, cred) \notin \{(p_1, cred_1), ... ,(p_n, cred_n)\}$, and $\huaA$ wins if $\cCredVrfy(p,cred,PK_{CA})=1$.
      	The unforgeability requires that, $\forall$ polynomial-time $\huaA$, $\Pr[\huaA $ wins$]$ is negligible.
      }
      \end{definition}
      The intuition of the strong unforgeability is that one cannot forge a valid anonymous credential for any given prefix, if she does not know the $SK$ corresponding to a certified by CA.
      }


  \smallskip
  The first one characterizes a special {\em accountability} requirement in anonymous authentication. It requires that no efficient adversary can authenticate two messages with a common-prefix without being linked, if using a same certificate. More generally, if an attacker corrupts $q$ users, she cannot authenticate $q+1$ messages sharing a common-prefix, without being noticed.  Formally, consider the following cryptographic game between a challenger $\cC$ and an attacker $\cA$:
  \begin{enumerate}
      \item $\cSetup$. The challenger $\cC$ runs the $\cSetup$ algorithm and obtains the master keys.
       \item $\cCertGen$ queries. The adversary $\cA$ submits $q$ public keys with different identities and obtains $q$ different certificates: $cert_1,\ldots,cert_q$.
      \item $\Auth$. The adversary $\cA$ chooses $q+1$ messages $p||m_1,\ldots,p||m_{q+1}$ sharing a common-prefix $p$ (with $|p|=\lambda$) and authenticates to the challenger $\cC$ by generating the corresponding attestations $\pi_1,\ldots,\pi_{q+1}$.
      \end{enumerate}
Adversary $\cA$ wins if all $q+1$ authentications pass the verification, and no pair of those authentications were linked.

      \theoremstyle{definition}
      \begin{definition}[Common-prefix-linkability]
      \label{def:linkable}

		For all probabilistic polynomial time algorithm $\huaA$, $\Pr[\huaA $ wins in the above game$]$ is negligible on the security parameter $\lambda$.
      \end{definition}


 \smallskip
 Next is the {\em anonymity} guarantee in normal cases. We would like to ensure the anonymity against any party, including the public, the registration authority, and the verifier who can ask for multiple (and potentially correlated) authentication queries. Also, our strong anonymity requires that no one can even tell whether a user is authenticating for different messages, if these messages have different prefixes.
 The basic requirement for anonymity is that no one can recognize the real identity from the authentication transcript. But our {\em unlinkability} requirement is strictly stronger, as if one can recognize identity, obviously, she can link two authentications by firstly recovering the actual identities.

 \smallskip
 To capture the unlinkability (among the authentications of different-prefix messages), we can imagine the most  stringent setting, where there are only two honest users in the system,  the adversary still cannot properly link any of them from a sequence of authentications. Formally, consider the following game between the challenger $\cC$ and adversary $\cA$.
       \begin{enumerate}
      \item $\cSetup$.  The adversary $\cA$  generates the master key pair.
       \item $\cCertGen$. The adversary $\cA$ runs the certificate generation procedure as a registration authority with the challenger. The challenger submits two public keys
       $pk_0,pk_1$ and the adversary generates the corresponding certificates for them $cert_0,cert_1$. $\cA$ can always generate certificates for public keys generated by herself.
      \item $\Auth$-queries. The adversary $\cA$ asks the challenger to serially use $(sk_0,pk_0,cert_0)$ and $(sk_1,pk_1,cert_1)$  to do a sequence of authentications on messages chosen by her.
      Also, the number $q$ of authentication queries is chosen by $\cA$. The adversary obtains $2q$ message-attestation pairs.
      \item $\Challenge$. The adversary $\cA$ chooses a new message $m^*$ which does not have a common prefix with any of the messages asked in the $\Auth$-queries, and asks the challenger to do one more authentication. $\cC$ picks a random bit $b$ and authenticates on $m^*$ using $sk_b,pk_b,cert_b$. After receiving the attestation $\pi_b$, $\cA$ outputs her guess $b'$.

      \end{enumerate}

The adversary wins if $b'=b$.

      \theoremstyle{definition}
      \begin{definition}[Anonymity]
      \label{def:unlinkable}
    \ignore{
      		The \emph{cross-prefix-unlinkability}  (anonymity) of \sameprefix-linkable anonymous credential can be defined with using the following challenger-adversary game:
      		(i) the challenger generates two different public-secret key pairs $(PK_0,SK_0)$ and $(PK_1,SK_1)$, and then registers at CA to obtain $cert_0$ binded to $PK_0$ and $cert_1$ binded to $PK_1$, also the challenger sends $PK_0$, $cert_0$, $PK_1$ and $cert_1$ to the adversary;
      		(ii) the adversary queries the challenger with a chosen prefix $p_i$, and the challenger computes and sends back $cred_0^i =\cCredGen(p_i,SK_0,PK_0,cert_0,PK_{CA})$ and $cred_1^i = \cCredGen(p_i,SK_1,PK_1,cert_1,PK_{CA})$, with specifying $cred_0^i$ is generated from $cert_0$ and $PK_0$, which will be done for $n$ times;
      		(iii) the adversary finally selects a set of prefix ${\bf P} = \{{p_1}',...,{p_m}'\}$ (s.t. ${\bf P} \cap \{p_1,...,p_n\} = \emptyset$ and $||{\bf P}|| \geq 2$) and send ${\bf P}$ to challenger;
      		for each ${p_j}' \in {\bf P}$, the challenger computes $cred_j = \cCredGen(p_j,SK_{b_j},PK_{b_j},cert_{b_j},PK_{CA})$ where $b_j \leftarrow \{0,1\}$, and sends $({p_j}',cert_{b_j})$ to the adversary; the adversary wins, if she outputs two prefixes ${p_k}' \in {\bf P}$ and ${p_l}' \in {\bf P}$, s.t. $b_k = b_l$ and ${p_k}'\neq {p_l}'$.
		}
      		An authentication scheme is unlinkable, if $\forall$ probabilistic polynomial-time algorithm $\huaA$, $|\Pr[\huaA $ wins in the above game$] - \frac{1}{2}|$ is negligible.\footnotemark
      \footnotetext{We remark that our definition of anonymity is strictly stronger than that definition of the event-oriented linkable group signature \cite{ASY06}, in which the RA can revoke user anonymity under certain conditions.}
      \end{definition} 


	\medskip
	\noindent{\bf Construction.} Now we proceed to construct such a primitive. Same as many anonymous authentication constructions, we will also use the zero-knowledge proof technique towards anonymity. In particular, we will leverage zk-SNARK to give an efficient construction. For the above concept of \sameprefix-linkable anonymous authentication, we need to further support the special accountability requirement. The idea is as follows, since the condition that ``breaks'' the linkability is common-prefix, thus the authentication will do a special treatment on the prefix. In particular, the authentication shows a tag committing to the prefix together with the user's secret key, and then presents zero-knowledge proof that such a tag is properly formed, i.e., computed by hashing the prefix and a secret key. To ensure other basic security notions, we will also compute the other tag that commits to the whole message. The user will further prove in zero-knowledge that the secret key corresponds to a certified public key.
	
	\smallskip
	Note that our main goal is to decentralize crowdsourcing, such a new anonymous authentication primitive could be further studied systematically in future works. Concretely, we present the detailed construction as follows:

      \begin{enumerate}
      \smallskip\item[-] $\cSetup(\lambda)$. This algorithm establishes the public parameters $PP$ that will be needed for the zk-SNARK system. Also, the algorithm generates a key pair $(msk,mpk)$ which is for a digital signature scheme.\footnote{To be more precise, the public parameter generation could be from another algorithm. For simplicity, we put it here. In the security game for anonymity, the adversary only generates $msk,mpk$, not the public parameter. }
      \smallskip\item[-]  $\cCertGen(msk,pk_i)$: This algorithm runs a signing algorithm on $pk_i$,\footnote{We  assume an external identification procedure that can check the actual identity bound to each public key, and the user key pairs are generated using common algorithms, e.g., for a digital signature. We ignore the details here.} and obtains a signature $\sigma_i$. It outputs $cert_i:=\sigma_i$.
      \smallskip\item[-] $\Auth(p||m,sk_i,pk_i,cert_i,PP)$: On inputing  a message $p||m$ having a prefix $p$.

      The algorithm first computes two tags (or interchangeably called headers later), $t_1=H(p,sk_i)$ and $t_2=H(p||m,sk_i)$, where $H$ is a secure hash function.
      Then, let $\vec{w}=(sk_i,pk_i,cert_i)$ represent the private witness, and $\vec{x}=(p||m,mpk)$ be all common knowledge, the algorithm runs zk-SNARK proving algorithm $\cProve(\vec{x},\vec{w},PP)$ for the following language $\huaL_{T}:= \{ t_1,t_2,\vec{x} = (p||m,mpk) \mid \exists \vec{w}=(sk_i,pk_i,cert_i)$ s.t. $\cCertVrfy(cert_i,pk_i,mpk)=1 \wedge pair(pk_i,sk_i)=1\wedge t_1 = H(p,sk_i) \wedge t_2=H(p||m,sk_i)\}$, {where the $\cCertVrfy$ algorithm checks the validity of the certificate using a signature verification, and $pair$ algorithm verifies whether two keys are a consistent public-secret key pair}. This prove algorithm yields a proof $\eta$ for the statement $\vec{x} \in \huaL$ (also for the proof-of-knowledge of $\vec{w}$).

      Finally, the algorithm outputs $\pi:=(t_1,t_2,\eta)$.

      \smallskip\item[-] $\Verify(p||m,\pi,mpk,PP)$: this algorithm runs the verifying algorithm of zk-SNARK $\cVerify$ on $\vec{x},\pi$ and $PP$, and outputs the decision bit $d\in\{0,1\}$.
      \smallskip\item[-] $\Link(m_1,\pi_1,m_2,\pi_2)$: On inputting two attestations $\pi_1:=(t_1^1,t_2^1,\eta_1)$ and $\pi_2:=(t_1^2,t_2^2,\eta_2)$, the algorithm simply checks $t_1^1\stackrel{?}{=}t_1^2$. If yes, output $1$; otherwise, output $0$. We also use $\Link(\pi_1,\pi_2)$ for short.
      \end{enumerate}

    \medskip
    \noindent{\bf Security analysis (sketch).} Here we briefly sketch the security analysis for the construction. As the scheme is of independent interests, we defer detailed analyses/reductions to an extended paper where the scientific behind will be formally studied.  

    \smallskip
  	Regarding {\em correctness}, it is trivial, because of the completeness of underlying SNARK.

	\smallskip
	Regarding {\em unforgeability}, we require an uncertified attacker cannot authenticate. The only transcripts can be seen by the adversary are headers and the zero-knowledge attestation. Headers include one generated by hashing the concatenation of $p||m,sk$. In order to provide a header, the attacker has to know the corresponding $sk$, as it can be extracted in the random oracle queries. Thus there are only two different ways for the attacker: (i) the attacker generates forges the certificate, which clearly violates the signature security; (ii) the attacker forges the attestation using an invalid certificate, which clearly violates the proof-of-knowledge of the zk-SNARK.
	
	
	
	\smallskip
	Regarding the {\em common-prefix-linkability}, it is also fairly straightforward, as the final authentication transcript contains a header  computed by $H(p,sk)$ which is an invariable for a common prefix $p$ using the same secret key $sk$.

  	
  	\smallskip
  	Regarding the {\em anonymity/unlinkability}, we require that after seeing a bunch of authentication transcripts from one user, the attacker cannot figure out whether a new authentication comes from the same user. This holds even if the attacker can be the registration authority that issues all the certificates. To see this, as the attacker will not be able to figure the value of the $sk$ from all public value, thus the headers/tags can be considered as random values. It follows that $H(p,sk)$ and a random value $r$ cannot be distinguished (similarly for $H(p||m,sk)$). More importantly, due to the zero-knowledge property of zk-SNARK, given $r$, a simulator can simulate a valid proof $\eta^*$ by controlling the common reference string of the zk-SNARK. That said, the public transcript $t_1,t_2,\eta$ can be simulated by $r_1,r_2,\eta^*$ where $r_1,r_2$ are uniform values, and $\eta^*$ is a simulated proof, all of which has nothing to do with the actual witness $sk$.

  	\smallskip
  	Summarizing the above intuitive analyses, we have the following theorem:

  	\begin{theorem}
  			Conditioned on that the hash function to be modeled as a random oracle and the zk-SNARK is zero-knowledge, the construction of the common-prefix-linkable anonymous authentication satisfies anonymity. Conditioned on the underlying digital signature scheme used is secure, and the zk-SNARK satisfies proof-of-knowledge, our construction of the common-prefix-linkable anonymous authentication will be unforgeable. It is also correct and common-prefix linkable.
  	\end{theorem}

\medskip
\subsection{The protocol of ZebraLancer}
\label{sec:protocol2}

Now we are ready to present a general protocol for a class of crowdsourcing tasks having proper quality-aware incentives mechanisms defined as in Section \ref{problem}. As zk-SNARK requires a setup phase, we consider that a setup algorithm generated the public parameters $PP$ for
this purpose, and published it as common knowledge.\footnote{This in practice can be done via a secure multiparty computation protocol \cite{BCG15} to eliminate potential backdoors. }
Our descriptions focuses on the application atop the open blockchain, and therefore omits details of sending messages through the underlying blockchain infrastructure. For example, ``one uses blockchain address $\alpha$ to send a message $m$ to the blockchain'' will represent that he broadcasts a blockchain transaction containing the message $m$, the public key associated to $\alpha$, and the signature properly generated under the corresponding secret key.

\begin{figure}[!h]
	\centering
	\includegraphics[width=8cm]{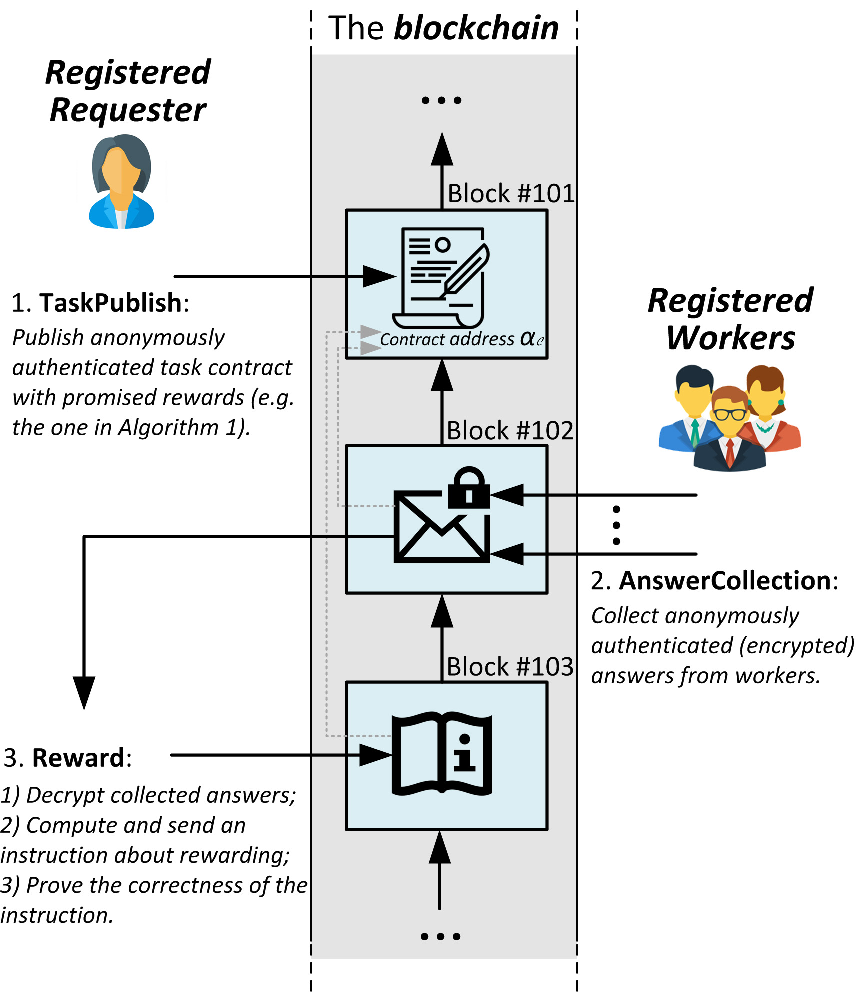}
	\caption{The schematic diagram of the protocol of {\system } for the quality-aware incentive mechanisms as proof-of-concept. Encrypted answers and the instruction of rewarding are pointing to the task published as a smart contract.}
	\label{fig:protcol}
\end{figure}

\begin{table}[!htbp]
		\renewcommand{\arraystretch}{1.3}
	\caption{Key notations related to the protocol of ZebraLancer}
	\label{table:protocol-notation}
	\begin{tabular}{m{1.3cm}||m{6.2cm}}
		\hline
		Notation & Represent  for \\
		\hline\hline
		$\Contract$ & The smart contract programmed for the crowdsourcing task to be published               \\[2pt]
		$\alpha_\Contract$ & The blockchain address of $\Contract$, which is also used as the prefix for authentication        \\[2pt]
		$\alpha_R$ & The one-time blockchain address used by the requester $R$ to anonymously interact with $\Contract$	\\[2pt]
		$\alpha_i$ & The one-time blockchain address used by the worker $W_i$ to anonymously interact with $\Contract$	\\[2pt]
		$(esk,epk)$ & The one-time public-secret key pair used in $\Contract$ for encrypting/decrypting crowd-shared answers	\\[2pt]
		$C_i$ &  The encrypted answer submitted to $\Contract$ by the worker $W_i$, i.e. the ciphertext of answer $A_i$	\\[2pt]
		
		$\overline{R}$ & The instruction sent from the requester to instruct the smart contract to reward each crowd-shared answer            	\\[2pt]
		$\pi_{reward}$ & The zk-SNARK proof generated to attest the correctness of the instruction 	of paying rewards \\[2pt]
		$\pi_i$ & The attestation sent from a user $i$ to anonymously authenticate a message having a prefix $\alpha_\Contract$            	\\[2pt]
		$PP$ & Public parameters for zero-knowledge proof, including the one for common-prefix-linkable anonymous authentication and the one for incentive mechanism\\[2pt]
		
		\hline
	\end{tabular}
\end{table}

\smallskip
Remark that here we let each worker/requester to generate a different blockchain address for each task (i.e. a {\em one-task-only} address) as a simple solution to avoid de-anonymization in the underlying blockchain.\footnote{Our anonymous protocol mainly focuses on the application layer such as the crowdsourcing functionality that is built on top of the blockchain infrastructure. If the underlying blockchain layer supports anonymous transaction, such as Zcash \cite{HBH17}, the worker and the requester can re-use account addresses. We further remark that the anonymity in network layer are out the scope of this paper, we may deploy our protocol on existing infrastructure such as Tor.}
For concrete instantiations of the underlying infrastructures, see the implementations in Section \ref{implementation}. 
We further remark that the protocol can be extended to private and anonymous auction-based incentives trivially (see Appendix B for a concrete example), although it mainly focuses on quality-aware ones.

\smallskip
\noindent \textbf{Protocol details.} As shown in Fig.{\ref{fig:protcol}}, the details of ZebraLancer protocol can be described as follows:

\begin{itemize}[itemsep=2pt]

	\smallskip\item $\mathsf {Register}$. {\em Everyone registers at RA to get a certificate bound to his/her unique ID, which is  done off-line only once for per each participant.}
	
	\smallskip A requester, having a unique ID denoted by $R$, creates a public-secret key pair $(pk_R,sk_R)$, and registers at the registration authority (RA) to obtain a certificate $cert_R$ binding $pk_R$ to $R$. Each worker, having a unique ID denoted by $W_i$, also generates his public and secret key pair $(pk_i,sk_i)$, and registers his public key at RA to obtain a certificate $cert_i$ binding $pk_i$ and $W_i$.

	\smallskip\item $\mathsf {TaskPublish}$. {\em A requester anonymously authenticates and publishes a task contract with a promised reward policy}.
	
	\smallskip When the requester  $R$ has a crowdsourcing task, she generates a {\em fresh} blockchain account address $\alpha_R$, 
	and a  key pair $(epk,esk)$ (which will be used for workers to encrypt submissions) for this task only.
	
	$R$ then prepares parameters $\cParam$, including the encryption key $epk$, the number of answers to collect (denoted by $n$), the deadline,
	the budget $\tau$,  the reward policy $\cR$, SNARK's public parameters $PP$, RA's public key $mpk$, and also $\pi_R =\Auth(\alpha_\Contract||\alpha_R,sk_R,pk_R,cert_R,PP)$.\footnote{We remark that the requester should authenticate on her blockchain address $\alpha_R$ along with the task contract, and workers will join the task only if the task contract is indeed sent from a blockchain address as same as the authenticated $\alpha_R$. So a malicious requester cannot ``authenticate'' a task by copying other valid authentications. In addition, each worker has to authenticate on his blockchain address $\alpha_i$ along with his answer submission as well. The task contract will check the submission is indeed sent from a blockchain address same to the authenticated $\alpha_i$. Otherwise, a malicious worker can launch free-riding through copying and re-sending authenticated submissions that have been broadcasted but not yet confirmed by a block. \label{fn:auth}}
	
	The requester then codes a smart contract $\Contract$ that contains all above information for her task. After compiling $\Contract$, she puts $\Contract$'s code and a transfer of the budget into a blockchain transaction, and uses the one-task-only address $\alpha_R$ to send the transaction into the blockchain network. When a block containing $\Contract$ is appended to the blockchain, $\Contract$ gets an immutable blockchain address $\alpha_\Contract$ to hold the budget and interact with anyone.
	\footnote{We emphasize that $\alpha_\Contract$ will be unique per each contract. In practice, $\alpha_\Contract$ can be computed via $H(\alpha_R||counter)$, where $H$ is a secure hash function, and $counter$ is governed by the blockchain to be increased by exact one for each contract created by the blockchain address $\alpha_R$. It's also clear that the requester $R$ can predicate $\alpha_\Contract$ before $\Contract$ is on-chain, such that she can compute $\pi_R$ off-line and let it be a parameter of contract $\Contract$.}
	
	
	See Algorithm \ref{contract1} below for a concrete example of task contract. (The important component of verifying zk-proofs is done by calling a library {\em libsnark.$\cVerify$} integrated into the blockchain infrastructure, and implementation details will be explained in Section \ref{implementation}).
	
	
	%

	\renewcommand{\algorithmicrequire}{\textbf{Input:}} 
	\renewcommand{\algorithmicensure}{\textbf{Output:}} 
	\SetKwProg{Fn}{function}{}{}
	\SetKwRepeat{Struct}{struct \{}{\}}
	\SetKwInOut{Require}{Require}

	\begin{algorithm}
		\label{contract1}\small
		\caption{Example using quality-aware incentive}
		\Require{This contract's address $\alpha_\Contract$; requester's one-time blockchain address $\alpha_R$; requester's authenticating attestation $\pi_R$; RA's public key $mpk$; budget $\tau$; public key $epk$ for encrypting answers; SNARK's public parameters $PP$; number of requested answers $n$; deadline of answering in unit of block $T_{\bm{A}}$; deadline of instructing reward in unit of block $T_{\bm{I}}$.}
		
		List keeping answers' ciphertexts, $\bm{C}\leftarrow\emptyset$\;
		Map of anonymous attestations and authenticated one-time blockchain addresses of workers, $\bm{W}\leftarrow\emptyset$\;

		\If {$getBalance(\alpha_\Contract) < \tau \vee \neg \Verify(\alpha_\Contract||\alpha_R,\pi_R,mpk,PP)$}
		{
			\textbf{goto 24} ; \Comment{Unidentified requester or no deposit.}
		}
		
		$timer_{\bm{A}}\leftarrow$ a timer expires after $T_{\bm{A}}$; \Comment{Collect answers}
		
		\While {$||\bm{C}|| < n \wedge timer_{\bm{A}}$ NOT expired}  
		{
			\If {$\alpha_i$ sends $\pi_i, C_i$}
			{
				\If {$ \neg \Link (\pi_i, \pi_R) \wedge  \forall  \pi_* \in  \bm{W}.keys() {\ }\neg\Link ( \pi_i, \pi_* ) \wedge \Verify(\alpha_\Contract||\alpha_i||C_i,\pi_i,mpk,PP)$ }
				{
					{$\bm{W}.add(\pi_i \rightarrow \alpha_i )$};
					{$\bm{C}.add(C_i)$};
				}
			}
		    \If {$\alpha_R$ sends $i, \pi_{fake}$} {
		    	\If {$libsnark.\cVerify((C_i,\sigma_i),\pi_{fake},PP)$}
		    	{
		    		{$\bm{W}.remove(\pi_i \rightarrow \alpha_i )$};
		    		{$\bm{C}.remove(C_i)$};
		    	}
		    }
		}
		
		$timer_{\bm{I}}\leftarrow$ a timer expires after $T_{\bm{I}}$;  \Comment{Wait instruction}
		
		\While {$timer_{\bm{R}}$ NOT expired}
		{
			\If {$\alpha_R$ sends $\overline{R}:=(R_1,\ldots,R_n)$ and $ \pi_{reward} $}
			{
				$\overline{P} \leftarrow (epk, \tau, C_1,\ldots,C_n)$\;
				\If {$libsnark.\cVerify((\overline{P},\overline{R}),\pi_{reward},PP)$}
				{
					\For{each $(\pi_i \rightarrow \alpha_i ) \in \bm{W}$}
					{
						$transfer(\alpha_\huaC,\alpha_i,R_i)$;
					}
					\textbf{goto 24};
				}
			}
			
		}
		$R \leftarrow \tau/||\bm{W}||$; \Comment{Reward all if no correct instruction}
		
		\For{each $(\pi_i \rightarrow \alpha_i) \in \bm{W}$}
		{
			$transfer(\alpha_\huaC,\alpha_i, R)$\;
		}
		$transfer(\alpha_\huaC$, $\alpha_R$, $getBalance(\alpha_\huaC))$; \Comment{Refund the else}

		\Fn{$getBalance (addr)$}{
			\textbf{return} the balance of $addr$ in the ledger;
		}
		\Fn{$transfer (src$, $dst$, $v)$}{
			\If {$getBalance (src) < v$}
			{
				\textbf{return} $false$;
			}
			$src$'s balance - $v$;
			$dst$'s balance + $v$;
			\textbf{return} $true$;
		}
		
		\Comment {The correctness and availability of this task contract is governed by the blockchain network; the contract program is driven by a ``discrete'' clock that increments with validating each newly proposed block}\\	
		\Comment {$libsnark.\cVerify$ is a library embedded in the runtime environment of smart contract such as EVM}\\
	\end{algorithm}

	\smallskip\item  $\mathsf{Answer Collection}$. {\em The contract collects anonymously authenticated encrypted answers from workers who didn't submit before.}
	
	\smallskip If a registered worker $W_i$ is interested in contributing, he first validates the contract content (e.g., checking the parameters), then generates a {\em one-time} blockchain address $\alpha_i$. 
	He signs his answer $A_i$ with using the private key associated with the blockchain address $\alpha_i$ to obtain a signature $\sigma_i$, then 
	 encrypts the signed answer $A_i||\sigma_i$ under the task's public key $epk$ to obtain ciphertext $C_i$. Note that $\sigma_i$ can be verified by the chain address $\alpha_i$, e.g., it is generated by the same signing algorithm to sign transactions.
	
	He then uses \sameprefix-linkable anonymous authentication scheme to generate an attestation $\pi_i =\Auth(\alpha_\Contract||\alpha_i||C_i,sk_i,pk_i,cert_i,PP)$.\footref{fn:auth}
Then he uses his {\em one-time} address $\alpha_i$ to send $C_i,\pi_i$ to the blockchain network (with a pointer to $\alpha_\Contract$, i.e. the unique address of the contract $\Contract$).

	Then, $\Contract$ runs $\Verify(\alpha_\Contract||\alpha_i||C_i,\pi_i,mpk,PP)$, and also executes $\Link(\pi_i,\pi_*)$ for each valid authentication attestation $\pi_*$ that was received before (including requester's, namely $\pi_R$). Such that, $\Contract$ can ensure  $C_i$ is the first submission of a registered worker.  For unauthenticated or double submissions, $\Contract$ simply drops it.\footnote{We remark that our protocol can be extended trivially to allow each worker to submit some $k$ answers in one task by modifying the checking condition programmed in the smart contract of crowdsourcing task.} 
	
	The contract $\Contract$ will keep on collecting answers, until it receives $n$ answers or the deadline (in unit of block) passes. It also records each address $\alpha_i$ that sends $C_i$. Remark that $\Link$ algorithm will be executed $O(n^2)$ times, but it is a simple equality check over a pair of hashes, such that the cost of running it for several times will be nearly nothing in practice.

	Along the way, the requester can keep on listening the blockchain, and decrypts each submission $C_i$ that is sent by $\alpha_i$ and accepted by the contract. Such that, the requester can obtain the signed answer $A_i||\sigma_i$ bound to $C_i$, and verifies the signature $\sigma_i$ using the blockchain address $\alpha_i$. If $A_i$ is not properly signed, the requester will generate a zero-knowledge proof $\pi_{fake}$ to convince the contract that  the plaintext bound to $C_i$ is not correctly signed message. Particularly, the proof is generated by zk-SNARK's $\cProve$ algorithm to attest the statement that $(C_i,epk,\alpha_i )\in$	
	$\{ {C_i,epk,\alpha_i} \mid \exists esk, s.t., verifying(\alpha_i,A_i,\sigma_i) \neq 1 \wedge A_i||\sigma_i=\Dec(esk,C_i) \wedge  pair(esk,epk)=1\} $, where $verifying$ is a function that outputs 1 iff the inputs correspond  properly signed message. Once the contract receives and validates such a proof $\pi_{fake}$ for a given submission from $\alpha_i$, it will remove $C_i$ and $\alpha_i$.
	Note that the above proof and verification are necessary to prevent the network adversaries who can delay others' submissions, and then copy and paste others' ciphertexts to reap rewards.

	\smallskip\item $\mathsf {Reward}$. {\em The requester computes and prove how to reward properly authenticated anonymous answers.}
	
\smallskip The requester $R$ keeps listening to the blockchain, and once $\Contract$ collects $n$ submissions, she retrieves and decrypts all of them to obtain the corresponding answers $A_1,\ldots, A_n$ (if there are not enough submissions when the deadline passes,
	the requester simply sets the remaining answers to be $\bot$ which has been considered by the incentive mechanism $\cR$).
	
	Next, the requester computes the reward for each answer $R_i=\cR(A_i;A_1,\ldots,A_n, \Budget)$ as specified by the policy codified in $\Contract$. More importantly, she generates a zero knowledge proof $\pi_{reward}$, with the secret key $esk$ as witness to attest the validity of the instruction. In particular, the proof is for the following NP-language	%
	$\huaL= \{\overline{R},\overline{P} \mid \exists esk$ s.t.  $\wedge_{j=1}^n A_j||\sigma_i = \Dec(esk,{C_j})\wedge_{j=1}^n R_j = \cR( A_j; A_1, \ldots ,A_n, \Budget) \wedge  pair(esk,epk)=1\} $, where $\overline{P}$ denotes $\cParam$ together with ciphertexts $C_1,\ldots,C_n$; while $\overline{R}:=(R_1,\ldots,R_n)$ is the instruction about how to reward each answer. After computing $\overline{R}$ and $\pi_{reward}$,
	$R$ puts them into a blockchain transaction, and still use her one-task-only blockchain address $\alpha_R$ to send the transaction to $\Contract$ (by using a pointer to $\alpha_\Contract$). This finishes the {\em outsource-then-prove} methodology.
	
	%
	Once a newly proposed block contains the reward instruction $\overline{R}$ and its attestation $\pi_{reward}$, the contract $\Contract$ first checks that they are indeed sent from $\alpha_R$ (by verifying the digital signature of the underlying blockchain transaction). Then it leverages SNARK's $\cVerify$ algorithm to verify the proof $\pi_{reward}$ regarding the correctness of $\overline{R}$. If the verification passes, it transfers each amount $R_i$ to each account $\alpha_i$, and refunds the remaining balance to $\alpha_R$. Otherwise, pause. If receiving no valid instruction after a predefined time (in unit of block), the contract simply transfers $\tau/n$ to each $\alpha_i$ as part of the policy $\cR$.
\end{itemize}
%


\medskip

\subsection{Analysis of the protocol}
\label{sec:analysis}

\noindent{\bf Correctness and efficiency.} It is clear to see that the requester will obtain data and the workers would receive the right amount of payments. If they all follow the protocol, under the conditions that (i) the blockchain can be modeled as an ideal public ledger, (ii) the underlying zk-SNARK is of completeness, (iii) the public key encryption is correct, and (iv) \sameprefix-linkable anonymous authentication satisfies correctness.
Regarding {\em efficiency}, we note the {\em on-chain} computation (and storage which are two of the major obstacles for applying blockchain in general) is actually very light, as the contract essentially only carries a verification step. Thanks to zk-SNARK, the verification can be efficiently executed by checking only a few pairing equalities; moreover, the special library can be dedicatedly optimized in various ways \cite{BCG13}.

\smallskip
\noindent \textbf{Security analysis (sketch).} 
We briefly discuss security here and defer the details to an extended version. The underlying primitives, including our \sameprefix-linkable anonymous authentication scheme, are well abstracted, which enable us to argue security in a modular way.

\smallskip
Regarding the {\em data confidentiality} of answers, all related public transcripts are simply the ciphertexts $C_1,\ldots,C_n$, and the zk-SNARK proof $\pi$. The ciphertexts are easily simulatable according to the semantic security of the public key encryption, and the proof $\pi$ can also be simulated without seeing the secret witness because of the {\em zero-knowledge} property.

\smallskip
Regarding the {\em anonymity}, an adversary has two ways to break it: (i) link  a worker/requester through his blockchain addresses; (ii) link answers/tasks of a worker/requester through his authenticating attestations.
The first case is trivial, simply because every worker/requester will interact with each task by a randomly generated one-task-only blockchain address (and the corresponding public key).
The second case is more involved, but the anonymity of workers and requesters can be derived through the anonymity of the \sameprefix-linkable anonymous authentication scheme.

\smallskip
Regarding the {\em security against a malicious requester}, a malicious requester has three chances to gain advantage: (i) deny the policy announced in \textsf {TaskPublish} phase; (ii) cheat in \textsf {Reward} phase; (iii) submit answers to intentionally downgrade others in \textsf {AnswerCollection} phase.
The first threat is prevented because the smart contract is public, and the requester cannot deny it once it is posted in the immutable blockchain. The second threat is prohibited by the soundness of the underlying zk-SNARK, since any incorrect instruction passing the verification in the smart contract, directly violates the proof-of-knowledge (i.e. the strong soundness). The last threat is simply handled the unforgeability and \sameprefix-linkability of our \sameprefix-linkable anonymous authentication scheme.

\smallskip
{\em Security against malicious workers} is straightforward, the only ways that malicious workers can cheat are: (i) submitting more than one answers in \textsf {AnswerCollection} phase; 
(ii) copying and pasting others' answers to earn reward; 
(iii) sending the contract a fake instruction in the name of requester in \textsf {Reward} phase; (iv) altering the policy specified in the contract. The first threat is simply handled by the common-prefix-linkability and unforgeability of \sameprefix-linkable anonymous authentication. The second threat can be approached by predicating the plaintexts of answers or copying the ciphertexts, but both are prevented: predicating plaintext is prevented due to the semantical security of public key encryption; copying the ciphertexts are prevented because the securities of digital signature and zk-SNARK. The third threat is simply handled by the security of digital signatures.  The last issue is trivial, because the blockchain security ensures the announced policy is immutable.

\begin{theorem}
\label{thm:basic}
The data confidentiality of our protocol holds, if the underlying public key encryption is semantically secure and the used zk-SNARK is of zero-knowledge.

The anonymity of our protocol for both workers and requesters will be satisfied, if the underlying \sameprefix-linkable anonymous authentication satisfies the anonymity defined in Definition \ref{def:unlinkable}, and the zk-SNARK is zero-knowledge.

Conditioned on that the blockchain infrastructure we rely on can be modeled as an ideal public ledger, the underlying \sameprefix-linkable anonymous authentication satisfies the unforgeability and the common-prefix-linkability, the zk-SNARK satisfies proof-of-knowledge and the digital signature in use is secure, our protocol satisfies: security against a malicious requester and security against malicious workers.

\end{theorem}

\ignore{
\begin{proof}
We defer the formal proof to an extended version.
\end{proof}
}


%

\medskip
\section{ZebraLancer: Implementation and Experimental Evaluation}\label{implementation}

We implement the protocol of {\system} atop Ethereum, and instantiate a series of typical image annotation tasks \cite{SZ15} with using it. Furthermore, we conduct experiments of these tasks in an Ethereum test net to evaluate the applicability.

\begin{figure}[!h]
	\centering
	\includegraphics[width=8.5cm]{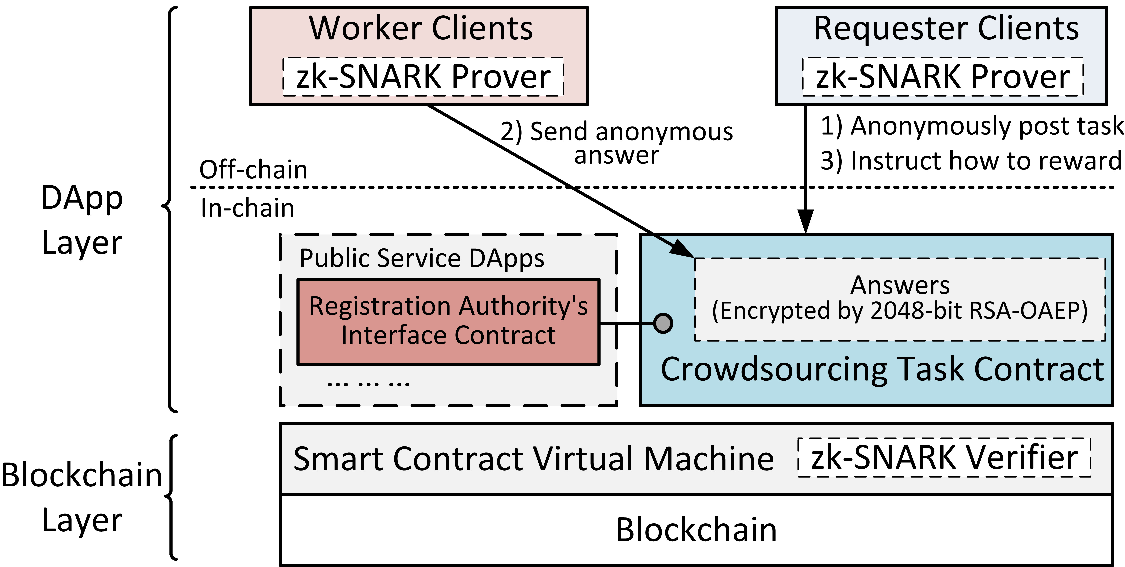}
	\caption{The system-level view of \system. Our Dapp layer can be built on top of an existing blockchain, e.g. the Ethereum Byzantium release \cite{Byzantium}.}
	\label{fig3}
\end{figure}

\smallskip
\noindent \textbf{System in a nutshell}. As shown in Fig.\ref{fig3}, the decentralized application (DApp) of our system is composed of an on-chain part and an off-chain part. The on-chain part consists of crowdsourcing task contracts and an interface contract of the registration authority (RA). The RA's contract simply posits the system's master public key as a common knowledge stored in the blockchain.
The off-chain part consists of requester clients and worker clients. These clients can be blockchain clients 
wrapped with functionalities required by our system. Specifically, a client of requester should codify a specific task with a given incentive mechanism and announces it as a smart contract.
Note that we, as the designers of the DApp, can provide contract templates to requesters for easier instantiation of incentive mechanisms, c.f. \cite{FN16}.
The clients further need an integrated zk-SNARK prover to produce the anonymous authenticating attestations; moreover, a requester client should also leverage SNARK prover to generate proofs attesting the correct execution of incentive policies.

\smallskip
We also compare \system{ }to some existing crowdsourcing systems in Table.\ref{table:comparison}. The security performance of our system overwhelms others, as it considers the most strict  fairness of exchange, user anonymity and data confidentiality, under that condition of minimum trust. For example, our design realizes the fair exchange without leaking data to a third-party information arbiter. And \system{ }also guarantees the strongest user anonymity that cannot be broken by any third-party (even the registration authority), while the anonymity of other systems can be broken by a third-party authority (or few colluded third-parties). Remark that the identities established via a registration service are required by all systems in Table.\ref{table:comparison}.

\begin{table}[!htbp]
\renewcommand{\arraystretch}{1.35}
	\caption{Comparison between our system \system{ }and some other crowdsourcing platforms}
	\label{table:comparison}
	\begin{tabular}{m{1.7cm}||m{0.7cm}m{0.85cm}m{0.85cm}m{0.85cm}m{0.85cm} }
		\hline
		& \scriptsize{Ours} & \scriptsize{MTurk} & \scriptsize{Dynamo}  & \scriptsize{SPPEAR}  & \scriptsize{CrowdBC}\\ 
		&  & \cite{MTurk} & \cite{SIB15}  & \cite{GGP16}  & \cite{LWY17}\\
		\hline\hline 	
		\scriptsize{Preventing false-reporting}    & $\surd$    & $\times$   & $\times$    & $\bigcirc$ & $\surd$ \\
		\scriptsize{Preventing free-riding}    & $\surd$    & $\bigcirc$  & $\bigcirc$   & $\bigcirc$  & $\surd$ \\
		\scriptsize{Data confidentiality}      & $\surd$    & $\times$    & $\times$     & $\times$   & $\times$ \\
		\scriptsize{User anonymity}     & $\surd$    & $\times$    & $\bigcirc$   & $\bigcirc$ & $\times$ \\ 
		\hline
	\end{tabular}
	\raggedright
	Note: $\surd$ denotes a  functionality realized without relying on any central trust except the established identities; $\bigcirc$ denotes a (partially) realized function by relying on a central trust (other than the registration authority); $\times$ denotes an unrealized feature. Note that the data confidentiality is marked as $\times$, if any third-party other than the requesters can access the submitted data.
\end{table}

\smallskip
\noindent{\bf Implementation challenges.} The main challenge of deploying smart contracts in general is that they can only support very light on-chain operations for both computing and storing.\footnote{We remark the communication overhead is not a serious worry, because: (i) a blockchain network such as Ethereum does not require fully meshed connections, i.e. requesters and workers can only connect a constant number of Ethereum peers; (ii) if necessary, requesters and workers can even run on top of so-called light-weight nodes, which eventually allows them receive and send messages only related to crowdsourcing tasks; (iii) even if there is a trusted arbiter facilitating incentive mechanisms, the only saving in communication is just an instruction about how to reward answers (and its attestation).} Our protocol actually has taken this into consideration. In particular, our on-chain computation only consists SNARK verifications, while the heavy computation of SNARK proofs are all done off the blockchain. Even still, building an efficient privacy-preserving DApp compatible with existing blockchain platform such as Ethereum is not straightforward. For instance, in order to allow smart contracts to call a zk-SNARK verification library, a contract of this library should be thrown into a block, but this library is a general purpose tool that can be too complex to be executed in the smart contract runtime environment, e.g. Ethereum Virtual Machine (EVM). Alternatively, we modify the the runtime environment of smart contracts, so that an optimized zk-SNARK verification library \cite{BCG13} is embedded in it as a primitive operation. Our modified Ethereum client is written in Java 1.8 with Spring framework, and is available at github.com/maxilbert/ethereumj.

\smallskip
We remark that Ethereum project recently integrated some new cryptographic primitives into EVM to enable SNARK verification as well \cite{Byzantium}, which ensures our DApp can essentially inherit all Ethereum users to maintain the blockchain infrastructure to govern the faithful execution of the smart contracts in our DApp.


\ignore{
\begin{figure}[!h]
\centering
\includegraphics[width=7.5cm]{fig5.eps}
\caption{Technical procedures of constructing the zk-SNARK in leak-free in-contract verification scheme.}
\label{fig2}
\end{figure}
}

\smallskip
\noindent
\textbf{Establishments of zk-SNARKs (off-line)}.
As the feasibility of \system{ }highly depends on the tininess of SNARK proofs and the efficiency of SNARK verifications, it becomes critical to establish necessary zk-SNARKs off-line. As formally discussed before, the authentication scheme and nearly all incentive mechanisms can be stated as some well-defined deterministic constraint relationships.
We first translate these mathematical statements into their corresponding boolean circuit satisfiability representations. Furthermore, we establish zk-SNARK for each boolean circuit, such that all required public parameters are generated. All the above steps are done off-line, as they are executed for only once when the system is launched. 
Note that the potential backdoors in these zk-SNARK public parameters could be further eliminated via an off-line protocol based on secure multi-party computation \cite{BCG15}. However, such an off-line setup is beyond the scope of showing our system feasibility.

\smallskip
\noindent
\textbf{An image annotation crowdsourcing task}.
To showcase the usability of our system, we implement a concrete crowdsourcing task of image annotation similar to the ones in \cite{SZ15}. The task is to solicit labels for an image which can later be used to train a learning machine. The task requests $n$ answers from $n$ workers, and can be considered as a multi-choice problem. Majority voting is used to estimate the ``truth''. An answer is seen as ``correct'', if it equals to the ``truth''. The reward amount of a worker is $\tau/n$ if he answers correctly, otherwise, he receives nothing.
 In our terminology, the reward $R_i:=\cR(A_i;A_1,\ldots,A_n,\tau)=\tau/n$, if $A_i$ equals the majority; otherwise, $R_i=0$. 
Following \cite{SZ15}, we implement and deploy 5 contracts in the test net to collect 3 answers, 5 answers, 7 answers, 9 answers and 11 answers from anonymous-yet-accountable workers, respectively.

\smallskip
The smart contracts are written in Solidity, a high-level scripting language translatable to smart contracts of Ethereum. We also modify Solidity compiler, such that a programmer can write a contract involving zk-SNARK verifications at high-level. We instantiate the encryption to be RSA-OAEP-2048, the DApp-layer hash function to be SHA-256, and the DApp-layer digital signature to be RSA signature. Moreover, for zk-SNARK, we choose the construction of {\em libsnark} from \cite{BCG13}.
We deploy a test network consisting of four PCs: three PCs are equipped with Intel Xeon E3-1220V2 CPU (PC-As), and the other one is equipped with Intel i7-4790 CPU (PC-B); all PCs have 16 GB main memory and have Ubuntu 14.04 LTS installed.
In the test net, a PC-A and a PC-B play the role of miners, and the other two PC-As only validate blocks (i.e. full nodes that do not mine). 
One full node plays the role of the requester, and anonymously publishes crowdsourcing tasks to the blockchain; and the other full node mimics workers, and sends each anonymously authenticated answer from a different blockchain address. Miners are only responsible to maintain the test net and do not involve in tasks.

\begin{table}[htpb]
\renewcommand{\arraystretch}{1.3}
\caption{Execution time of in-contract zk-SNARK verifications.}
\label{snark}
\begin{tabular}{c||ccc||ccc}
\hline
\multirow{2}{*}{\scriptsize{Verification for}}
& \multicolumn{3}{c||}{\scriptsize{Operands Length}}
& \multirow{2}{*}
{\begin{tabular}[c]{@{}c@{}}\scriptsize{Time@}\\\scriptsize{PC-A}\end{tabular}}
& \multirow{2}{*}
{\begin{tabular}[c]{@{}c@{}}\scriptsize{Time@}\\\scriptsize{PC-B}\end{tabular}}
\\ \cline{2-4}
& \scriptsize{Proof} & \scriptsize{Key}  & \scriptsize{Inputs}  & \\ \hline\hline
\scriptsize{Our authentication}     & \scriptsize{729B} & \scriptsize{1.2KB}  & \scriptsize{1.5KB} & \scriptsize{10.9$ms$} & \scriptsize{6.2$ms$} \\[3pt]
\scriptsize{Majority (3-Worker)} & \scriptsize{729B} & \scriptsize{16.0KB} & \scriptsize{3.4KB} & \scriptsize{15.5$ms$} & \scriptsize{9.1$ms$}\\[3pt]
\scriptsize{Majority (5-Worker)} & \scriptsize{730B} &  \scriptsize{21.6KB} &  \scriptsize{4.7KB} & \scriptsize{16.3$ms$} & \scriptsize{9.8$ms$}\\[3pt]
\scriptsize{Majority (7-Worker)} & \scriptsize{731B} & \scriptsize{27.3KB} & \scriptsize{6.0KB} & \scriptsize{17.0$ms$} & \scriptsize{10.3$ms$}\\[3pt]
\scriptsize{Majority (9-Worker)} & \scriptsize{729B} &  \scriptsize{32.9KB} &  \scriptsize{7.3KB} & \scriptsize{17.5$ms$} & \scriptsize{12.1$ms$}\\[3pt]
\scriptsize{Majority (11-Worker)} & \scriptsize{730B}  & \scriptsize{38.6KB} & \scriptsize{8.6KB} & \scriptsize{17.9$ms$} & \scriptsize{13.1$ms$}\\[3pt]
\hline
\end{tabular}
\\
Note: the sizes of proofs, keys and inputs are their bytes after being serialized in unicode.
\end{table}

\smallskip
\noindent
\textbf{Performance evaluation}. As the main bottleneck is the on-chain computation of the smart contract, we first measure the time cost and the spatial cost of miners, regarding the executions of zk-SNARK verifications used in the above annotation tasks. These zk-SNARKs are established for \sameprefix-linkable anonymous authentications and incentive mechanisms, respectively. The results of time cost are listed in Table \ref{snark}. It is clear that zk-SNARK verifications in our system can be efficiently executed in respect of verification time. Moreover, our experiment results also reveal that the spatial cost of zk-SNARK verifications is constant and tiny at both types of PCs (exactly 17$MB$ main memory). Also, the required on-chain storage for the task contracts is at the acceptable magnitude of kilobyte.
Therefore, the on-chain performance of the system can be clearly practical, considering both time and space.

At the off-chain side of requester, we consider reformation of the statements of attesting correct payments to enhance the off-chain efficiency, as the inherent heavy NP-reductions of generating zk-SNARK proofs could be prohibitively expensive for the requester without careful optimizations. Briefly speaking, we reform $\huaL= \{\overline{R},\overline{P} \mid \exists esk$ s.t.  $\wedge_{j=1}^n A_j||\sigma_i = \Dec(esk,{C_j})\wedge_{j=1}^n R_j = \cR( A_j; A_1, \ldots ,A_n, \Budget) \wedge  pair(esk,epk)=1\} $, and translate it into $\huaL'= \{\overline{R},\overline{P} \mid \exists esk$ s.t.  $\wedge_{j=1}^n {C_j} = \Enc(epk,A_j||\sigma_i)\wedge_{j=1}^n R_j = \cR( A_j; A_1, \ldots ,A_n, \Budget) \wedge  pair(esk,epk)=1\}$, because $epk$, as the exponent of modular exponentiation operations, is significantly smaller than $esk$ in RSA, and therefore the latter language corresponds a more compact arithmetic circuit and essentially brings more efficient proving. The proving cost at the requester end therefore becomes acceptable; for example, to prove the majority of 11 workers' binary voting, the requester spends 56 GB memory+swap and less than 2 hours, which was hundreds of gigabytes and about one day without such the optimization \cite{LTW18}.

\begin{figure}[!h]
\centering
\includegraphics[width=6.5cm]{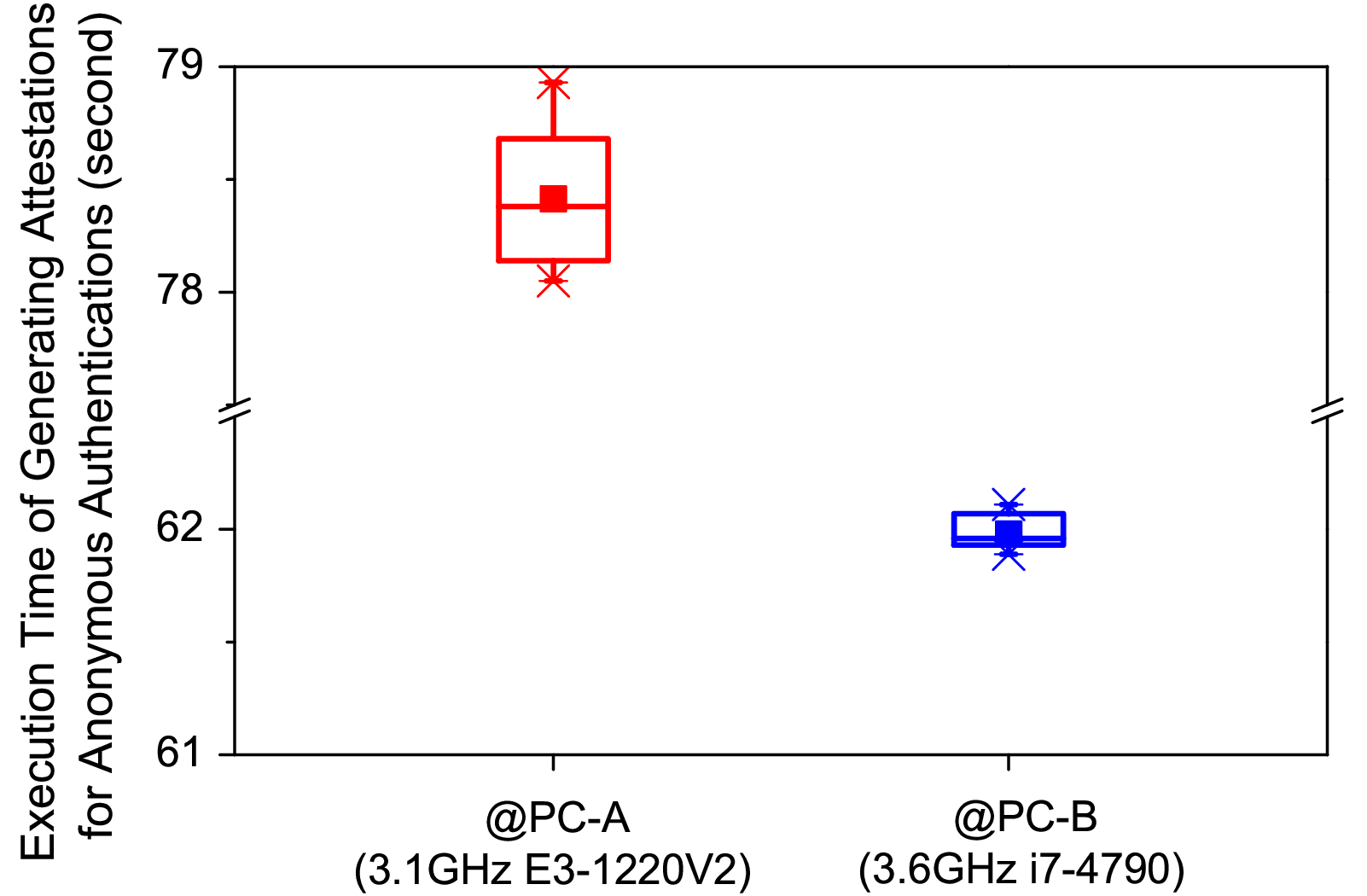}
\caption{The time of generating \sameprefix-linkable anonymous authentications in two PCs. The box plot is derived from 12 different experiments.}
\label{proof}
\end{figure}

\smallskip
We also measure the off-chain cost of anonymity, if one uses the \sameprefix-linkable anonymous authentication. We measure the running time of generating the authenticating attestations at PCs.
As shown in Fig.\ref{proof}, our experiment results clarify that
about 78 seconds are spent on generating an anonymous attestation with using PC-A (3.1 GHz CPU). In PC-B (3.6GHz CPU), the running time can be shortened to about 62 seconds.
Those are not ideal, but acceptable by the anonymity-sensitive workers. We remark that our protocol can be trivially extended to support non-anonymous mode, in case that one gives up the anonymity privilege: s/he can generate a public-private key pair (for digital signatures), and then registers the public key at RA to receive a certificate bound to the public key; to authenticate, s/he can simply show the certified public key, the certificate, along with a message properly signed under the corresponding secret key, which essentially costs nearly nothing regarding the computational efficiency.

\medskip

\medskip
\section{Extension for Auction-Based Incentives}

\smallskip
The pseudocode in Algorithm 2 showcases  how to extend the ZebraLancer protocol to instantiate a private and anonymous crowdsourcing task using an auction-based incentive mechanism to pre-select workers. The main difference is using auction to pre-select workers before they submit data (line 6-12). As shown in the algorithm, the workers can signed their bids (with using their private keys associated with the blockchain addresses), and then encrypt these signed bids under requester's public key. The ciphertexts can be sent to the task contract (line 7-9). Then, the requester is allowed to convince the contract that the plaintext bound to a submission is not properly signed by the corresponding worker  (line 10-12). Moreover, the requester is then incentivized to leverage the {\em outsource-then-prove} method to prove the result of auction (i.e. the selected workers) to the task contract (line 13-18). After the workers are selected, they can submit answers to the contract (line 20-25).

\renewcommand{\algorithmicrequire}{\textbf{Input:}} 
\renewcommand{\algorithmicensure}{\textbf{Output:}} 
\SetKwInOut{Require}{Require}

\begin{algorithm}
	\label{contract2}\small
	\caption{Example of auction-based incentive}
	\Require{This contract's address $\alpha_\Contract$; requester's one-time blockchain address $\alpha_R$; requester's authenticating attestation $\pi_R$; RA's public key $mpk$; budget $\tau$;  public key $epk$ for encrypting bids and answers; SNARK's public parameters $PP$;
		deadline of bidding in unit of block $T_{\bm{B}}$; deadline of instructing auction result in unit of block $T_{\bm{I}}$; deadline of answering in unit of block $T_{\bm{A}}$; deadline of instructing auction result in unit of block $T_{\bm{I}}$.}
	
	List keeping the ciphertexts of bids, $\bm{B}\leftarrow\emptyset$\;
	Map of anonymous attestations and authenticated one-time blockchain addresses of workers, $\bm{W}\leftarrow\emptyset$\;

	\If {$getBalance(\alpha_\Contract) < \tau \vee \neg \Verify(\alpha_\Contract||\alpha_R,\pi_R,mpk,PP)$}
	{
		\textbf{goto 26} ; \Comment{Unidentified requester or no deposit.}
	}
	
	$timer_{\bm{B}}\leftarrow$ a timer expires in $T_{\bm{B}}$; \Comment{Collect secret bids.}
	
	\While {$timer_{\bm{B}}$ NOT expired}
	{
		\If {$\alpha_i$ sends $\pi_i$, and his encrypted bid $B_i$}
		{
			\If {$ \neg \Link (\pi_i, \pi_R) \wedge  \forall  \pi_* \in  \bm{W}.keys() {\ }\neg\Link ( \pi_i, \pi_* ) \wedge \Verify(\alpha_\Contract||\alpha_i,\pi_i,mpk,PP)$ }
			{
				{$\bm{W}.add( W_i:= (\pi_i \rightarrow \alpha_i) )$};
				{$\bm{B}.add(B_i)$};
			}
		}
		\If {$\alpha_R$ sends $i, \pi_{fake}$} {
			\If {$libsnark.\cVerify((B_i,\sigma_i),\pi_{fake},PP)$}
			{
				{$\bm{W}.remove(\pi_i \rightarrow \alpha_i )$};
				{$\bm{B}.remove(B_i)$};
			}
	}
	}
	
	$timer_{\bm{I}}\leftarrow$ a timer expires after $T_{\bm{I}}$;  \Comment{Wait instruction}\
	
	\While {$timer_{\bm{I}}$ NOT expired}
	{
		\If {$\alpha_R$ sends selected workers $\overline{S}:=(W_{1'},\ldots,W_{n'})$ and $ \pi_{auction} $}
		{
			$\overline{P} \leftarrow (epk, \tau, B_1,\ldots,B_{||\bm{B}||}, W_1,\ldots,W_{||\bm{W}||})$\;
			\If {$libsnark.\cVerify((\overline{P},\overline{S}),\pi_{auction},PP)$}
			{
				\textbf{goto 20};
			}
		}
		
	}
	$\overline{S}=\bm{W}$; \Comment{All workers are ``selected'' if no instruction.}

	$timer_{\bm{A}}\leftarrow$ a timer expires after $T_{\bm{A}}$;  \Comment{Collect answers}\
	
	\While {$||\overline{S}|| \neq 0  \wedge timer_{\bm{A}}$ NOT expired}
	{
		\If {$\alpha_i$ sends encrypted answer $C_i$}
		{
			\If {$ (* \rightarrow \alpha_i) \in \overline{S}$ }
			{
				$transfer(\alpha_\huaC,\alpha_i, B_i)$\;
				$\overline{S}.remove((* \rightarrow \alpha_i))$;
			}
		}
	}

	$transfer(\alpha_\huaC$, $\alpha_R$, $getBalance(\alpha_\huaC))$; \Comment{Refund the else}


\end{algorithm}

\smallskip
The main motivation of the above ``private'' bidding process is to prevent a malicious worker from learning the bidding strategies of honest workers through exploring all historic bids stored in the open blockchain, which is an implicit requirement of most auction-based incentive mechanisms (mainly for truthfulness). For example, when the requester is using an incentive mechanism (e.g. a variant of reverse Vickrey auction \cite{ZLM14}) to make workers give their truthful bids, if a malicious worker can well estimate the distribution of all other bids, then an optimized bid, which is usually untruthful, can be explored by him to break the requester's auction mechanism. Note that the public parameter of the zk-SNARK for verifying the result of the auction (i.e. pre-selected workers) is established off-line and published as common knowledge, such that it can be referred in the contract by $PP$.

\smallskip
Remark  that the private auction atop blockchain is an interesting orthogonal problem \cite{auction} per se.  Nevertheless, \system{} is really general, and can be adapted for a very broad variety of incentive mechanisms including auctions, although we didn't intentionally focus on private auctions. 

\smallskip
It is also clear that our \sameprefix-linkable anonymous authentication scheme can  be straightforwardly used to make the auction-based extension to be anonymous-yet-accountable, as the multiple bidding of each worker is prevented by its subtle common-prefix linkability. To further allow a worker submit $k$ bids in a single task, we can instantiate a counter in the smart contract to track the number of bids from each worker, because of the \sameprefix-linkability property.

\medskip

\section{Conclusion \& Open Problems}\label{sec6}

\system{\ }can facilitate a large variety of incentive mechanisms to realize the fair exchange between the crowd-shared data and their corresponding rewards, without the involvement of any third-party arbiter. Moreover, it shows the practicability to resolve two natural tensions in the use-case of the decentralized crowdsourcing atop open blockchain: one between the data confidentiality and the blockchain transparency, and the other one between the participants' anonymity and the their accountability.

\smallskip
Along the way, we put forth a new anonymous authentication scheme. Besides the strong anonymity that cannot be revoked by any authority, it also supports  a delicate linkability only for messages that share the common prefix and are authenticated by the same user. A concrete construction of the scheme is proposed, and it shows the compatibility to real-world blockchain infrastructure. The delicate linkability of the scheme is subtly different from the state-of-the-art of anonymous-yet-accountable authentication schemes \cite{Cha83,CFN90,CL01,XY04,TFS04,CHK06,ASY06}, and we envision the scheme might be of independent interests.
We also develop a general {\em outsource-then-prove} technique to use smart contracts in a privacy-preserving way. This technique can further extend the scope of applications atop some existing privacy-preserving blockchain infrastructures such as \cite{HBH17,Byzantium,KMS16}.


\medskip
\noindent{\bf Open problems}. Since this work is the first attempt of decentralizing data crowdsourcing  atop the real-world blockchain in a privacy-preserving way, the area remains largely unexplored. Here we name a few open questions, and we defer solutions to them in our future work. First, there are many incentive mechanisms using reputation systems, can we further extend our implementations to support those incentives? Second, as the current smart contract technology is at an infant stage and can only allow very tiny on-chain storage, can we further optimize our implementations for particular crowdsourcing tasks to assist more large-scale tasks, e.g. to collect annotations for millions of images (i.e. the scale of ImageNet dataset)? Third, our anonymous protocol currently either relies on the underlying blockchain to support anonymous transaction, or requires workers/requesters use one-time blockchain account to submit data and receive reward. Can we design a (DApp-layer) protocol to solve the drawbacks? Last but not least, our protocol relies on a trusted registration authority (RA) to establish identities. 
Although such a trusted RA could be a reasonable assumption (in view of real-world experiences), it is more tempting to develop an alternative methodology to remove the third-party RA without sacrificing securities. For example, can we adapt the successful invention of proof-of-work to build up a crowdsourcing framework from literally ``zero'' trust without any established identity?


\medskip
\bibliographystyle{IEEEtran}
{\footnotesize
\bibliography{IEEEabrv,reference}
}

\end{document}